\documentclass[11pt, a4paper]{scrartcl}

\usepackage[utf8]{inputenc}	
\usepackage[english]{babel}
\usepackage[T1]{fontenc}
\usepackage{lmodern} 

\usepackage{float}
\usepackage[top = 2.5cm, bottom = 3cm, left = 4cm, right = 4cm]{geometry} 

\usepackage{amsmath,mathtools,amsthm} 
\usepackage{dsfont} 
\usepackage{amssymb}
\usepackage{comment}
\usepackage{authblk} 

\usepackage[ruled,vlined]{algorithm2e}
\usepackage{makecell}
\usepackage{diagbox}
\usepackage{paralist}
\usepackage{multirow}
\usepackage{booktabs}
\usepackage{rotating}

\RestyleAlgo{ruled}
\LinesNumbered
\SetKwComment{Comment}{// }{}

\usepackage[printonlyused,withpage]{acronym}
\makeatletter
\AtBeginDocument{%
  \renewcommand*{\AC@hyperlink}[2]{%
    \begingroup
      \hypersetup{hidelinks}%
      \hyperlink{#1}{#2}%
    \endgroup
  }%
}
\makeatother
\usepackage{subcaption}
\usepackage{graphicx}
\usepackage[table,dvipsnames]{xcolor}
\usepackage[colorlinks=true,
			hyperindex, 
		    linkcolor = NavyBlue,
		    anchorcolor = hyperrefblue,
		    citecolor = Green,
		    filecolor = hyperrefblue,
		    urlcolor = violet,
			breaklinks=true]{hyperref}

\usepackage{tikz}
\usepackage[numbers,sort&compress]{natbib}

  \providecommand{\pre}{Phys.\ Rev.\ E}
  \providecommand{\prl}{Phys.\ Rev.\ Lett.}

\usepackage[capitalize]{cleveref}

\newcommand{\RR}{\mathbb{R}}

\renewcommand{\vec}[1]{\mathbf{#1}} 

\DeclarePairedDelimiterX{\abs}[1]{\lvert}{\rvert}{%
	\ifblank{#1}{\,\cdot\,}{#1}
}   

\DeclarePairedDelimiterX\norm[1]\lVert\rVert{%
	\ifblank{#1}{\,\cdot\,}{#1}
}   

\DeclarePairedDelimiterX{\iiiNorm}[1]{\lvert}{\rvert}{%
	\delimsize\lvert\delimsize\lvert#1\delimsize\rvert\delimsize\rvert%
}

\DeclarePairedDelimiterXPP\snorm[1]{}\lVert\rVert{_\infty}{\ifblank{#1}{\,\cdot\,}{#1}}   

\DeclarePairedDelimiterXPP\twonorm[1]{}\lVert\rVert{_2}{\ifblank{#1}{\,\cdot\,}{#1}}   

\DeclarePairedDelimiterXPP\trnorm[1]{}\lVert\rVert{_1}{\ifblank{#1}{\,\cdot\,}{#1}}   

\DeclarePairedDelimiterXPP\fnorm[1]{}\lVert\rVert{_{\fro}}{\ifblank{#1}{\,\cdot\,}{#1}}   

\DeclarePairedDelimiterXPP\dnorm[1]{}\lVert\rVert{_\diamond}{\ifblank{#1}{\,\cdot\,}{#1}}   

\DeclarePairedDelimiterXPP\cbnorm[1]{}\lVert\rVert{_\mathrm{cb}}{\ifblank{#1}{\,\cdot\,}{#1}}   
\DeclarePairedDelimiterXPP\onenorm[1]{}\lVert\rVert{_{1\rightarrow 1}}{\ifblank{#1}{\,\cdot\,}{#1}}   
\DeclarePairedDelimiterXPP\ddnorm[1]{}\lVert\rVert{_{\diamond\rightarrow \diamond}}{\ifblank{#1}{\,\cdot\,}{#1}}   
\DeclarePairedDelimiterXPP\ssnorm[1]{}\lVert\rVert{_{\infty\rightarrow\infty}}{\ifblank{#1}{\,\cdot\,}{#1}}   

\DeclarePairedDelimiterX\Set[1]\{\}{%
	
	#1
}

\DeclarePairedDelimiterX\innerp[2]{\langle}{\rangle}{%
	\ifblank{#1}{\,\cdot\,}{#1} , \ifblank{#2}{\,\cdot\,}{#2}%
}

\DeclarePairedDelimiterX\braket[2]{\langle}{\rangle}%
{#1\kern0.15ex\delimsize\vert\kern0.15ex\mathopen{}#2}

\DeclarePairedDelimiterX\ketbra[2]{\vert}{\vert}%
{#1\kern0.15ex\delimsize\rangle\delimsize\langle\kern0.15ex\mathopen{}#2}

\DeclarePairedDelimiterX\sandwich[3]{\langle}{\rangle}%
{#1\,\delimsize\vert\kern0.15ex\mathopen{}#2\kern0.15ex\delimsize\vert\kern0.15ex\mathopen{}#3}

\DeclarePairedDelimiterX\obraket[2]{(}{)}%
{#1\kern0.15ex\delimsize\vert\kern0.15ex\mathopen{}#2}

\DeclarePairedDelimiterX\oketbra[2]{\vert}{\vert}%
{#1\kern0.15ex\delimsize)\delimsize(\kern0.15ex\mathopen{}#2}

\DeclarePairedDelimiterX\osandwich[3]{(}{)}%
{#1\,\delimsize\vert\kern0.15ex\mathopen{}#2\kern0.15ex\delimsize\vert\kern0.15ex\mathopen{}#3}

\definecolor{martin}{rgb}{0,.4,1}

\definecolor{lightgray1}{gray}{0.95}
\definecolor{lightgray2}{gray}{0.9}

\newtoggle{showmodif}      
\toggletrue{showmodif}     
\togglefalse{showmodif}  

\newtoggle{showmodiff}      
\toggletrue{showmodiff}     
\togglefalse{showmodiff}  

\newcommand{\modiff}[1]{%
  \iftoggle{showmodiff}{\textcolor{red}{#1}}{#1}%
}

\title{A comprehensive benchmark of an Ising machine on the Max-Cut problem}

\author[1]{Salwa Shaglel\thanks{corresponding author}\thanks{salwa.shaglel@tuhh.de}}
\author[2]{Markus Kirsch\thanks{markus.kirsch@fujitsu.com}}
\author[1]{Marten Winkler}
\author[2]{Christian Münch}
\author[2]{Stefan Walter}
\author[2]{Fritz Schinkel}
\author[1]{Martin Kliesch}

\affil[1]{Institute for Quantum Inspired and Quantum Optimization, 
          Hamburg University of Technology,
          Blohmstraße 15, 21079 Hamburg, Germany}

\affil[2]{Fujitsu Germany GmbH, 
    Mies-van-der-Rohe-Straße 8, 80807 Munich, Germany}

\date{}    

\begin{document}

\maketitle

\hypersetup{
    pdftitle = {A comprehensive benchmark of an Ising machine on the Max-Cut problem},
    pdfauthor = {Salwa Shaglel, 
                 Markus Kirsch, 
                 Marten Winkler, 
                 Christian Münch,
                 Stefan Walter, 
                 Fritz Schinkel, 
                 Martin Kliesch}
    pdfsubject = {Quantum inspired computing},
    pdfkeywords = {quantum, inspired, computing, computer, combinatorial, optimization, benchmarking,  meta-heuristics,  algorithm, comparison, quadratic, unconstrained, binary, digital, annealer}
}
\begin{abstract}
QUBO formulations of combinatorial optimization problems allow for solving them using various quantum heuristics. While large-scale quantum computations are currently still out of reach, we can already numerically test such QUBO formulations on a perhaps surprisingly large scale. 

In this work, we benchmark Fujitsu's Digital Annealer (DA) on the Max-Cut problem, which captures the main complexity of the QUBO problem. We make a comprehensive benchmark against leading other heuristic algorithms on graphs with up to 53,000 variables by focusing on the wall-clock time. Moreover, we compare the DA performance against published performance results of the D-Wave hybrid quantum-classical annealer and the recently proposed QIS3 heuristic. Based on performance statistics for over 2,000 graphs from the MQLib, we find that the DA yields competitive results. We hope that this benchmark demonstrates the extent to which large QUBO instances can be heuristically solved today, yielding consistent results across different solvers. 
\end{abstract}

\tableofcontents

\section{Introduction}
\label{sec:intro}
Combinatorial optimization problems are ubiquitous in many applications, including vehicle routing problems, product assembly optimization, finance portfolio optimization, RNA folding problems, machine learning, and numerous other domains \cite{Glover_Kochenberger_Du_2022}. 
Often, they can be formulated as \acf{QUBO} problems \cite{Lucas2014, Glover2018, Ratke2021, Glover_Kochenberger_Du_2022}, which naturally fit certain quantum computing approaches. 
Since quantum hardware is not sufficiently mature yet to solve these problems at practically relevant scales, often classical \ac{QUBO} solvers are used to solve large \ac{QUBO} problems.
\modiff{Therefore, the \ac{QUBO} solvers are often called \emph{quantum-inspired} even though the underlying basic methods, such as simulated annealing and parallel tempering, predate the quantum computing efforts.}

Indeed, in the current absence of sufficiently powerful quantum hardware, it is difficult to assess the performance of a quantum heuristic. 
At the same time, it is important to single out problems where the \ac{QUBO} approach is particularly promising. 
This information can be essential for improving the quantum-readiness of those problems. 
It is highly desirable for \ac{QUBO} formulations of practical use-cases to be tested today at as large a scale as possible to gain insight into how potential quantum computing approaches might solve relevant \ac{QUBO} problems.
Current quantum hardware can operate with an order of 1000 noisy qubits. 
In contrast, dedicated \ac{QUBO} solvers can deal with tens of thousands of bits without suffering from noise and severe connectivity constraints typically encountered in quantum hardware.

The \ac{QUBO} problem is NP-hard \cite{punnen2022qubo}, and as such, it is not expected to be solvable in polynomial time, even with quantum computers \cite{aaronson2008limits}. This fundamental limitation has driven recent research toward the development of heuristic algorithms and the design of specialized classical \modiff{and quantum}
hardware. 
These approaches intentionally trade off optimality guarantees in favor of scalability and speed, aiming to tackle intractable problems in practice. 
Notable examples include the simulated bifurcation machine by Toshiba \cite{sqbm,Aatsumura21ScalingOutIsing}, the quantum annealer by D-Wave \cite{dwave2025}, Hitachi's Momentum Machines \cite{Okuyama19BinaryOptimizationBy}, NTT's coherent Ising machines \cite{Inagaki16CoherentIsingMachine,Honjo21_100k-spin}, and the digital annealer by Fujitsu \cite{Fujitsu}. 
The goal of this work is to shed light on the current capabilities of such a so-called \emph{Ising machine} \cite{Mohseni2022}--the \acf{DA}. 
The \ac{DA} is a
CMOS-based ASIC designed to efficiently execute an enhanced simulated annealing algorithm.
\modiff{By some it is referred to as `quantum-inspired' because it solves a class of problems that is native to certain quantum computing approaches.}

Among the many QUBO-encoded problems, the \acf{Max-Cut} problem stands out as a natural NP-hard problem that admits a straightforward \ac{QUBO} formulation. 
This ease of formulation, in addition to its relevance across graph theory and combinatorial optimization, has made \ac{Max-Cut} a popular benchmark for testing and comparing optimization solvers. 
Goemans and Williamson \cite{goemans1995improved} introduced a polynomial-time randomized approximation algorithm based on semidefinite programming, achieving a worst-case performance guarantee (approximation ratio) of at least $ \approx 0.87856$ on \ac{Max-Cut} (with nonnegative weights).
Assuming the unique games conjecture and $\mathsf{P}\neq\mathsf{NP}$, that is the best value that can be achieved in polynomial time \cite{Khot07OptimalInapproximabilityResults}.
In contrast to approximation algorithms, as of 2018 \cite{Dunning2018}, over 95 heuristics have been proposed for \ac{Max-Cut}. 
These heuristics prioritize computational efficiency and practical performance over guarantees and are widely employed in large-scale applications where exact or approximate algorithms become infeasible.

In this work, we benchmark and compare different algorithms and hardware. In order to have a useful and meaningful benchmark, it must be set up, executed, and reported on as transparently as possible. 
Without such transparency, a benchmark can lack rigor, fairness, or even become misleading \cite{Phillipson2025, Dunning2018}.
In particular, besides the general benchmark setup and hardware specifications, the following information should be provided:

\begin{enumerate}
    \item\label{item:BenchmarkSet}
    \textbf{Benchmark set}: \hypertarget{1a}{(1a)} Convincing reasons for the selection of the chosen benchmark set. 
    \hypertarget{1b}{(1b)} The set should reflect the diversity and complexity of real-world problem distributions. In particular, \hypertarget{1c}{(1c)} it should not be chosen due to solver-dependent considerations.
    \item\label{item:Runtimes}
    \textbf{Runtimes}: \hypertarget{2a}{(2a)} Convincing arguments for the chosen runtime limits, and \hypertarget{2b}{(2b)} deviations from those limits; \hypertarget{2c}{(2c)} tuning times and to what extent they have been included in the runtime. Not fully including tuning times in the runtime should be convincingly justified.
    \item\label{item:PerformanceDetails} \textbf{Performance details}: \hypertarget{3a}{(3a)} Convergence speeds, 
    \hypertarget{3b}{(3b)} number of trials, 
    \hypertarget{3c}{(3c)} parallelization of the algorithm, and 
    \hypertarget{3d}{(3d)} the obtained solutions (not just the objective values) so that independent verification is possible.
\end{enumerate}

Our study follows all of these desiderata by conducting a systematic and fair enough comparison of the second- and third-generation \ac{DA}, referred to as \ac{DA}v2 and \ac{DA}v3, respectively, on a broad and diverse set of \ac{Max-Cut} instances. Throughout the methodology and results sections, we refer back to these points to indicate where and how each criterion has been addressed.

First, we compare the \ac{DA}’s performance against the known best classical heuristics from \acs{MQLib}. 
To this end, we build on a benchmark of 37 heuristics \cite{Dunning2018} by selecting the most powerful ones for our comparison. 
More specifically, we choose those heuristics that demonstrated the best performance across different combinations of instance size and density. 
Our benchmark covers a total of 2,125 \ac{Max-Cut} instances. 
To promote fairness and comparability, we adopt the methodology of Ref.~\cite{Dunning2018}, which assigns a tailored time limit to each instance based on its computational difficulty and accounts for expected hardware improvements over time.

To complement this benchmark, we compare the performance of \ac{DA}v3 with D-Wave’s hybrid quantum-classical solver on their selection of 45 \ac{Max-Cut} instances \cite{dwave_hybrid_solver_2020}, as well as with the recent quantum-inspired metaheuristic QIS3 solver \cite{yang2025} on their selection of 16 instances.
In both comparisons, we shed light on \ac{DA}'s rapid convergence properties. A summary of the benchmarks performed in this work is presented in \cref{table:summary_work}.
\begin{table}[!t]
    \centering \resizebox{\textwidth}{!}{
    \begin{tabular}{|ll|l|l|l|l|l|l|} \hline
       \multicolumn{2}{|l|}{}   &
       \multicolumn{2}{c|}{\textbf{\acs{MQLib} heuristics} \cite{Dunning2018}} & \multicolumn{2}{c|}{\textbf{D-Wave \acs{HS}} \cite{dwave_hybrid_solver_2020}} & \multicolumn{2}{c|}{\textbf{QIS3} \cite{yang2025}}\\ \cline{3-8} 
     \multicolumn{2}{|l|}{\multirow{-2}{*}{\textbf{\diagbox[height=\rotheadsize +1.9\line]
         {\raisebox{0.5ex}{Spec.\hspace{2.2ex}}}{\raisebox{-0.5ex}{\llap{Solvers}}}}}} &
       \textbf{\ac{DA}v2} & \textbf{\ac{DA}v3} & \multicolumn{1}{l}{\textbf{\ac{DA}v3}} &  & \textbf{\ac{DA}v2}  & \textbf{\ac{DA}v3}   \\ \hline
        \multirow{2}{*}{Instances} & amount & 2,044  &  819  & \multicolumn{1}{l}{45} & &  14 & 16  \\ 
        \cline{2-8} & size & [200 --  8,176] & [2,048 --  53,130] & \multicolumn{1}{l}{[2,000 --  10,000]} &  & [800 --  8,000] & [800 --  10,000]\\ 
        \cline{2-8} &  density & [0.00032 -- 1] & [0.0001 -- 0.995]  & \multicolumn{1}{l}{[0.0004 --  0.97]} &  & [0.0005 -- 0.06] & [0.0004 -- 0.06] \\  
        \multicolumn{2}{|l|}{Library}  & \multicolumn{2}{c|}{\acs{MQLib} \cite{mqlib}} & \multicolumn{2}{c|}{\acs{MQLib}} &  \multicolumn{2}{c|}{G-set \cite{ye2003gset}} \\ 
        \multicolumn{2}{|l|}{Time limit}  & \multicolumn{2}{c|}{instance-specific} &  \multicolumn{2}{c|}{20 minutes} & \multicolumn{2}{c|}{10 seconds} \\ \hline
    \end{tabular}}
    \caption{Summary of benchmarks considered in this work. 
    For the benchmark on the \acs{MQLib} heuristics, we restrict our benchmark to those instances that can be \modiff{  reliably solved within the assigned time and that fit the chip (For transparency, the raw data of the unanalyzed instances are also available on GitHub~\cite{shaglel2025dabenchmark})}.
    For the benchmarks against D-Wave’s \acs{HS} and QIS3, we follow their original selections and use their provided data. 
    }
    \label{table:summary_work}
\end{table}

\subsection{Previous work}
\label{sec:previous_work}
Quantum computing has introduced several paradigms aimed at solving combinatorial optimization problems such as \ac{Max-Cut}. One of the most studied gate-based quantum approaches is the \ac{QAOA}, introduced by \citet{Farhi2014AQA}. \ac{QAOA} provides approximate solutions for combinatorial optimization problems using a variational circuit. It has been specifically evaluated on 2-regular and 3-regular \ac{Max-Cut} instances with single-layer depth, achieving a theoretical approximation ratio of at least 0.6924. This ratio is expected to improve with increasing circuit depth, as proven in \cite{Zhou2018QuantumAO}. For a detailed overview of QAOA and its extensions, see \cite{Blekos2023ARO}.
Despite its potential, \ac{QAOA} remains in its early stages of development. Demonstrating consistent quantum advantage with \ac{QAOA} is currently out of reach and remains an active area of research due to several ongoing challenges \cite{Blekos2023ARO}. 
These include: 
finding tractable problems for quantum computers \cite{Farhi14BoundedOccurrenceConstraint,Barak15BeatingTheRandom}, and the complexity of parameter optimization as the circuit depth increases due to barren plateaus \cite{McClean_2018},
local minima \cite{Bittel2021TrainingVQ} 
and hardness of finding hyperparameters \cite{BittelOptimizingTheDepth}. 
Moreover, the impact of hardware noise \cite{Guerreschi_2019, Wang_2021} and overheads due to noise reduction \cite{PRXQuantum.1.020312,cai2023} pose additional challenges. 

In parallel to gate-based quantum computation, a distinct approach to solving the \ac{Max-Cut} problem has emerged through quantum annealing and related Ising machines.
Quantum annealers, such as those built by D-Wave Systems, are a prominent hardware realization of this approach \cite{Hamerly_2019}.
The state-of-the-art quantum annealer is the D-Wave Advantage2 quantum processing unit, which contains over 4,400 superconducting qubits and utilizes the Zephyr topology with degree-20 connectivity \cite{dwave2025}. This marks a substantial improvement over the earlier Pegasus-based Advantage system \cite{boothby2020nextgeneration}.
In a recent benchmark reported in D-Wave's white paper \cite{dwave2025}, both Advantage and Advantage2 were used to solve the largest 3D-lattice spin glasses, which can be embeddable in their respective topologies---a $12 \times 12 \times 12$ cube with 1,650 variables and 4,461 edges. The Advantage2 system consistently returned better-quality solutions under identical annealing times and, in many cases, outperformed the Advantage system by several orders of magnitude in speed. 
Nonetheless, despite these technological advances, current quantum annealers face several critical limitations. These include restricted qubit counts, limited connectivity, and short coherence times \cite{Hauke_2020}. As the number of qubits increases, so does the complexity of coupler layout, introducing more noise---especially when many couplers are redundant for a given problem instance \cite{Gonzalez_Calaza_2021, Willsch_2022}. These constraints pose serious challenges for the scalability and universality of quantum annealers.

Given the current limitations of quantum hardware, quantum-inspired approaches offer a practical alternative for tackling combinatorial optimization problems.
Building upon the previously reviewed hardware considerations, recent studies have sought to empirically benchmark the performance of emerging quantum and quantum-inspired annealing platforms. Notably, Fujitsu's \ac{DA} has been evaluated against both classical heuristics and quantum annealing systems.

\citet{matsubara2020} provide an evaluation of \ac{DA}v2, using 65 instances from the G-set benchmark suite \cite{ye2003gset}.
These instances, which span sparse graphs with edge densities between 0.0002 and 0.06 and variable counts ranging from 800 to 8000, were compared against results obtained using IBM's CPLEX solver by Ref.~\cite{ikuta2015maximumcut} and a classical Multiple Operators heuristic \cite{ma2015multiple}. 
The study finds that \ac{DA}v2 achieves the best cut values for 62 out of the 65 instances and delivers the shortest runtime for 52 of them, underscoring its efficiency on sparse yet moderately sized \ac{Max-Cut} problems.
Our work improves upon this study by addressing several of its limitations, specifically those related to \hyperlink{1b}{(1b)}, \hyperlink{2a}{(2a)}, \hyperlink{3a}{(3a)}, and \hyperlink{3d}{(3d)} from the benchmarking criteria list.

A more nuanced comparison is provided by \citet{huang2022}, who evaluate both the D-Wave quantum annealer and Fujitsu's \ac{DA}v2 across 10 \ac{Max-Cut} instances ranging from 543 to 5430 variables and constrained to Pegasus- or Chimera-like graphs to conform to the architecture of D-Wave’s quantum processing unit. Their results show that Pegasus-based and Chimera-based quantum annealers (used in earlier D-Wave systems) perform well on smaller or sparsely connected graphs but show diminishing results on larger or denser instances in comparison to \ac{DA}. To further examine the dependence of solver's performance on graph structure, \citet{huang2022} generated 32 \ac{Max-Cut} instances, each with 145 variables and average degrees ranging from 1 to 140. While these instances were still more or less tailored to D-Wave’s quantum processing unit connectivity, the \ac{DA} demonstrated improved performance compared to the quantum annealer.
This study does not meet several important criteria from the list, particularly \hyperlink{1b}{(1b)}, \hyperlink{1c}{(1c)}, and \hyperlink{3d}{(3d)}. Although the study explicitly discusses hyperparameter tuning, it excludes the tuning process from the reported runtime, thereby not fully including \hyperlink{2c}{(2c)}.

Complementing the comparative studies discussed earlier, Oshiyama and Ohzeki \cite{Oshiyama_2022} conducted an extensive benchmark involving multiple advanced solvers: D-Wave’s \ac{HS}, Toshiba’s Simulated Bifurcation Machine (SBM), Fujitsu’s \ac{DA}v2, and simulated annealing. Their evaluation focused on the \ac{Max-Cut} problem using 45 instances (chosen by D-Wave \cite{dwave_hybrid_solver_2020}) from the \ac{MQLib}, restricted to problem sizes up to 8000 variables. Their findings indicate that \ac{DA}v2 outperformed the other solvers on large instances, although it did yield slightly inferior results on a few specific cases. Aggregated, D-Wave's \ac{HS} achieved the best results on 22 out of the 45 instances, followed by \ac{DA}v2 (20), SBM (16), and simulated annealing (7). When broken down by problem size, D-Wave's \ac{HS} led on small instances, while \ac{DA}v2 dominated on medium and large ones. In terms of edge density, D-Wave's \ac{HS} was most effective on sparse graphs, whereas \ac{DA}v2 performed best on graphs with medium to high connectivity.
Our study extends this comparison by benchmarking \ac{DA}v3 against D-Wave’s \ac{HS}, while addressing important gaps in notably \hyperlink{1b}{(1b)}, \hyperlink{2a}{(2a)}, \hyperlink{3a}{(3a)}, and \hyperlink{3d}{(3d)}.

Further evidence of \ac{DA}’s competitive performance is presented in a separate benchmarking study by \citet{cseker2022}. This study compares \ac{DA}v2 to three exact solvers—GUROBI \cite{gurobi}, CPLEX \cite{ikuta2015maximumcut}, and SCIP \cite{Achterberg2009SCIPSC}—as well as two classical heuristics: BURER2002 \cite{Burer2002RankTwoRH} and PAL2004bMTS2 \cite{Palubeckis2004MultistartTS} as implemented by \citet{Dunning2018}. Using a sample of 260 \ac{Max-Cut} instances with up to 8000 vertices, and runtime limits of 60 and 120 seconds, the results show that \ac{DA}v2 consistently yields solution quality and runtimes that are better
or competitive, further reinforcing its promise for practical, time-constrained optimization. 
However, there is still potential to include, in particular, the criteria \hyperlink{2a}{(2a)}, \hyperlink{2b}{(2b)}, and \hyperlink{3d}{(3d)}. While some aspects of \hyperlink{3a}{(3a)} are briefly mentioned, we provide concrete evidence to support it.

\citet{Jiang2024} benchmark three hardware-based annealers---D-Wave Advantage, Fujitsu’s \ac{DA}v3, and Quantix GPUA---alongside a classical solver from Meta-Analytics, across several NP-hard problems including \ac{Max-Cut}.
For \ac{Max-Cut}, they evaluate 10 instances with 2000 vertices at varying densities. Despite the limited sample, \ac{DA}v3 consistently finds the \ac{Max-Cut} across all instances and yields better results than other hardware solvers in both solution quality and runtime across all problems considered. 
Several important aspects are not covered by this study, including \hyperlink{1a}{(1a)}, \hyperlink{1b}{(1b)}, \hyperlink{2a}{(2a)}, \hyperlink{3a}{(3a)}, and \hyperlink{3d}{(3d)}.

Beyond \ac{Max-Cut}, Fujitsu’s \ac{DA} has been applied successfully to various NP-hard combinatorial optimization problems reformulated as \ac{QUBO} models. Studies have explored the \ac{DA}’s performance on graph problems such as 3-Regular 3-XORSAT \cite{kowalsky2022}, quadratic assignment \cite{cseker2022, huang2022, matsubara2020}, minimum-cut \cite{matsubara2020}, 3-SAT \cite{chris2023}, NAE 3-SAT \cite{Oshiyama_2022}, number and graph partitioning \cite{Kao2023}, and minimum vertex cover \cite{huang2022}. Additionally, the \ac{DA} has been used effectively in scientific domains including transport robot scheduling \cite{leib2023}, molecular structure optimization \cite{lee2024}, magnetic phase discovery \cite{jha2021}, and 2-D magnetic array design for energy harvesting and planar motors \cite{maruo2020}. These studies generally report favorable solution quality and time-to-solution for the \ac{DA}. Moreover, \ac{QUBO} modeling facilitates integration into multi-phase solution strategies. For instance, \citet{dornemann2025} employ the \ac{DA} in the final phase of a three-step approach to the capacitated vehicle routing problem with time windows, using it to solve a set partition problem and obtain a feasible global solution.

\subsection{Our contribution}
We aim to answer the question of how \ac{QUBO} solver performances compare from a user perspective. 
Therefore, we focus on the wall-clock time and compare the obtained objective values. 

While several previous studies already provide some benchmarks, we put an emphasis on following all the desired benchmarking standards on 
\ref{item:BenchmarkSet}.\ the benchmark set selection, 
\ref{item:Runtimes}.\ the runtime choices, and 
\ref{item:PerformanceDetails}.\ reporting  performance details (see \cref{sec:intro}). 

Our main study includes a diverse subset of heterogeneous instances from \ac{MQLib}, representing the largest test suite among all works reviewed in \cref{sec:previous_work} \hyperlink{1b}{(1b)}.
Following \citet{Dunning2018}, we obtain instance-specific baseline time limits determined as the convergence time of a simple greedy local search heuristic \hyperlink{2a}{(2a)}. Each heuristic (\acp{DA} and \ac{MQLib}) is then run individually on the same server to avoid interference from background load or parallel processes. 
By establishing these updated time limits, rather than reusing those from Ref.~\cite{Dunning2018}, we take the hardware specifications into account and assign proper time limits that capture the computational complexity of instances. This avoids both overestimating and underestimating solver performance.

This sets the stage for our first selection of instances from \ac{MQLib}. We first exclude all instances with time limits below 0.25 seconds, as these are considered too easy and prone to the solver's runtime overshooting. The second filtering criterion is based on size (number of variables).
A single \ac{DAU} supports instances up to 8,192 variables, so for \ac{DA}v2 we consider instances up to this size.
In contrast, \ac{DA}v3 has an additional software layer capable of handling instances of up to 100,000 variables. However, we restrict our analysis to instances with at least 2,048 variables, as we observed significant runtime overshooting on smaller instances, making the comparison less meaningful \hyperlink{1a}{(1a)}. Importantly, the remaining instances are not selected based on the \ac{DA}'s considerations (e.g., connectivity, edge-weight distribution or type, structural features) to avoid bias toward instances known to favor \ac{DA} performance \hyperlink{1c}{(1c)}. 
The details of this benchmark set are presented in \cref{table:summary_work}.

In addition, we report actual runtimes rather than just the time limits and explicitly analyze deviations from these limits \hyperlink{2b}{(2b)}. 
We also include in the total runtime the overhead from the \ac{DA}'s automatic internal hyperparameter tuning, and avoid any further manual parameter tuning \hyperlink{2c}{(2c)}.

We emphasize that the computational architectures we consider differ significantly in nature. 
The \ac{DA} is a highly parallelized heuristic \hyperlink{3c}{(3c)}, whereas the classical baseline heuristics from \ac{MQLib} are single-threaded solvers. 
The aim of this work is not to modify the solvers but to evaluate them as provided, enabling us to observe performance differences across computational architectures. Indeed, our results indicate that for certain instances, a single-threaded classical heuristic solver can outperform a parallelized dedicated one. 
Nevertheless, our direct comparison in terms of wall-clock time naturally favors a parallel architecture. Parallelized or GPU-based classical solvers (e.g., multithreaded simulated annealing, parallel tempering, or modern GPU heuristics) are not included in the present study. A review of available implementations did not yield candidates that are publicly available, well-documented, and reproducible solvers to be integrated into a large-scale and reproducible benchmark as we aim in this work. This consideration was the main motivation for including additional complementary benchmarks with D-Wave’s \ac{HS} and QIS3.

For the benchmarks involving D-Wave’s \ac{HS} and QIS3, we use the benchmark sets and runtime limits provided in the papers \cite{dwave_hybrid_solver_2020, yang2025}.
Therefore, we deviate from some of our benchmarking criteria, particularly \hyperlink{1a}{(1a)}, \hyperlink{1b}{(1b)}, potentially \hyperlink{1c}{(1c)}, and \hyperlink{2a}{(2a)}, as these aspects are shaped by how the original solver studies were conducted. 
However, we still ensure the criteria \hyperlink{2b}{(2b)}, \hyperlink{2c}{(2c)}, \hyperlink{3a}{(3a)}, \hyperlink{3b}{(3b)}, and \hyperlink{3d}{(3d)}, thereby providing as rigorous and transparent a comparison as currently feasible. We particularly analyze the convergence behavior of \ac{DA}v3~\hyperlink{3a}{(3a)}.

For all solvers executed in this work, we perform 5 runs with different seeds and base the benchmark on the best objective value among them \hyperlink{3b}{(3b)}. The best-of-5 summary reflects an application-driven setting in which users can afford only a small number of restarts and select the best outcome. In addition, whenever applicable, we provide statistical inference using majority-based paired run-level significance tests to rigorously evaluate whether observed differences between solvers are statistically meaningful rather than due to random variation. Additional reporting of robustness statistics across seeds (e.g., median and standard deviation) and the solution configurations (binary variable vectors) for all instances involved in this study are made available in our Git repository \cite{shaglel2025dabenchmark} \hyperlink{3d}{(3d)}.

The rest of this paper is organized as follows. \Cref{sec:background} provides the necessary background, including an overview of \ac{QUBO}, the \ac{Max-Cut} formulation, and a fundamental introduction to the \ac{DA}. The detailed methodology of our benchmarks against classical heuristics, D-Wave's \ac{HS}, and QIS3 is described in \cref{sec:methods}. Finally, we present and discuss the performance results in \cref{sec:results}.

\section{Background}
\label{sec:background}
This section provides a general introduction to \ac{QUBO} problems, the \ac{Max-Cut} problem, and the \ac{DA}, along with the basic definitions and notations used throughout this work.

\subsection{QUBO Problems}
\label{subsec:qubo}
Many combinatorial optimization problems can be formulated as a \ac{QUBO} problem given as
\begin{equation}
	\begin{aligned}
		&\mathrm{minimize
        } \hspace{0.2cm} \vec{x}^\top Q \vec{x}   = \sum_{i \leq j}^n q_{ij} x_i x_j, \\
        & \mathrm{subject \, \, to} \hspace{0.2cm} \vec{x} \in \{0,1\}^n, 
	\end{aligned}
	\label{eq:qubo_formula}
\end{equation}
where $Q \in \RR^{n \times n}$ is an upper-triangular matrix with diagonal elements $q_{ii}$ corresponding to linear terms (since $x_{i}x_{i} = x_i$ for $x_i \in \{0,1\}$), while the off-diagonal elements $q_{ij}$ for $i \neq j$ correspond to quadratic terms. The goal is to find an optimal solution configuration $\vec{x}^* \in \{0,1\}^n$ that minimizes the quadratic objective function~\eqref{eq:qubo_formula}. 
Constrained models can also be reformulated as \ac{QUBO} problems by introducing penalty terms into the objective function, instead of explicitly imposing constraints. These penalty terms are typically constructed to be zero for feasible configurations and positive for infeasible ones. 
A penalty prefactor is typically included to control the cost of constraint violation, and its value must be appropriately chosen to ensure the resulting solution is both optimal and feasible \cite{Glover2018}.

In fact, \cref{eq:qubo_formula} is
directly related to finding the ground state of the Ising Hamiltonian $- \sum_{i=1}^n h_i s_i - \sum_{i>j}^n J_{ij} s_i s_j$, where spin variables $\vec{s} \in \{-1,1\}^n$ relate to $\vec{x} \in \{0,1\}^n$ by $s_i = 2x_i - 1$. In this formulation, the coefficients $J_{ij}$ represent the interaction strength between spins $s_i$ and $s_j$, and $h_i$ corresponds to the strength of the external magnetic field acting on spin $s_i$. This connection underlines the naming of many hardware solvers designed for such problems as Ising machines \cite{Mohseni2022}.

The \ac{QUBO} problem is NP-hard. Therefore, all \ac{QUBO} solvers are expected to exhibit a worst-case runtime that scales exponentially with the size of the problem $n$, also for quantum solvers. That is, general large instances are expected to be computationally intractable. 
However, for commonly chosen instances and tolerable approximation errors, the scaling might be more favorable.
Many NP-hard problems have already been formulated in the \ac{QUBO} form, with 112 such formulations listed in Ratke's blog~\cite{Ratke2021}.

\subsubsection*{The \texorpdfstring{\ac{Max-Cut}}{Max-Cut} problem}
\label{subsubsec:maxcut}
Given a weighted undirected graph with a vertex set $V$, edge set $E$, and symmetric weight matrix $W$, a cut in this graph is a set of edges with exactly one endpoint in $S \subset V$:
\begin{equation}
	 \delta(S) = \{ \{v_i,v_j\} \in E | \ v_i \in S \text{ and } v_j \in \bar{S}  \},
\end{equation}
where $\bar{S} = V \setminus S$ is the complement set of $S$. The \ac{Max-Cut} problem is to find a cut of $V$ into two disjoint subsets $S \subset V$ and $\bar{S}$ with maximum sum of edge weights $w_{ij}$ between them, i.e., $ f(S) = \sum_{\{v_i,v_j\} \in \delta(S)} w_{ij}$ is maximized.
We will refer to $f(S)$ as the \emph{objective value} or the \emph{cut value}, which will be used interchangeably throughout this paper.
We call an instance \emph{unweighted} if all edges have weight one, i.e., $w_{ij} = 1$ if $\{v_i,v_j\} \in E$ and $w_{ij} = 0$ otherwise. 
In this case, the objective is to maximize the number of edges in the cut, and thus the cut value is then given by $f(S) = |\delta(S)|$.

The \ac{Max-Cut} problem is directly reducible to a \ac{QUBO} problem with little overhead, making it a representative problem that captures the main computational challenges of \ac{QUBO} problems. 
As such, any \ac{Max-Cut} problem on a graph can be expressed as in \cref{eq:qubo_formula} with $|V|$ binary variables.
Every vertex $v_i \in V$ in a graph corresponds to a binary variable $x_i \in \{0,1\}$ in the \ac{QUBO} problem. We assign the variable $x_i$ to $1$ if $v_i \in S$ and $0$ otherwise. For an edge between $v_i$ and $v_j$ to be in the cut $\delta(S)$, it must thus hold $x_i \neq x_j$.
The maximized cut objective has quadratic coefficients $q_{ij} = -2 w_{ij}$ (for $i<j$) and $q_{ii} = \sum_{j \neq i} w_{ij}$ and can be written as
\begin{equation}
	\label{eq:maxcut}
	f(\vec{x}) = \sum_{\{i,j\} \in E} w_{ij} (x_i + x_j - 2 x_i x_j),
\end{equation}
where $(x_i + x_j - 2 x_i x_j)$ evaluates to $1$ if $x_i$ and $ x_j$ are different and $0$ otherwise. In particular, maximizing \cref{eq:maxcut} is equivalent to solving the \ac{Max-Cut} problem. 
Since, by convention, Ising machines minimize rather than maximize, the equivalent problem is to minimize $-f(\vec{x})$. The corresponding minimization formulation in \cref{eq:qubo_formula} therefore uses the negatives of the coefficients above.

The equivalence also implies that any \ac{QUBO} instance can be formulated as a \ac{Max-Cut} instance in a graph with $|V|+1$ vertices. 
While coefficients of quadratic terms in the \ac{QUBO} problem correspond directly to edge weights between vertices in a graph, linear terms do not have a natural graph representation. 
However, they can be embedded into the graph structure by introducing an auxiliary vertex $x_0$, connecting it to all other vertices in the graph, and fixing its value to 1. This allows linear terms to be interpreted as additional edge weights incident on $x_0$. 
A quadratic objective $\vec{x}^\top Q \vec{x}$ can be represented as a cut value, up to an additive constant, by assigning $w_{ij} = -\frac{1}{2} q_{ij} $ (for $i<j$) and $w_{0j} = \sum_{ \substack{i > 0 \\ i \neq j}} w_{ij} - q_{jj}$. To obtain a Max-Cut instance equivalent to the minimization problem in \cref{eq:qubo_formula}, this construction is applied to $-Q$.

\subsection{The Digital Annealer}
\label{subsec:digital_annealer}
The \ac{DA} is a technology specifically designed to heuristically address combinatorial optimization problems formulated in \ac{QUBO} form. 
Its core component, the \ac{DAU}, is an application-specific integrated circuit (ASIC) constructed using CMOS technology, capable of directly realizing \ac{QUBO} problems with up to 8,192 binary variables.
The \ac{DAU} employs a simulated annealing variant as a basic search algorithm, and is directly compatible with any connectivity. 
This flexibility allows running the \ac{DA} natively on arbitrary graphs, in contrast to hardware architectures that require native graph topologies or special graph embeddings.

The simulated annealing algorithm is based on a Markov chain Monte Carlo (MCMC) method, namely the Metropolis-Hastings algorithm.
It begins with an initial bit-string configuration $\vec{x} = (x_1, x_2, \cdots, x_n)$ at a high temperature $T$, which is a hyperparameter of the algorithm. 
At each update step, a candidate configuration $\vec{x}^\prime \in \{0,1\}^n$ is generated by flipping a randomly chosen variable $x_i \in \vec{x}$ for some $i \in \Set{1,2,3,…,n}$ and the energy difference $ \Delta E = E(\vec{x}^\prime) - E(\vec{x})$ is computed. 
The new configuration is accepted with probability
\begin{equation}
	 p = \min \{\exp(- \Delta E / T ),1\}. 
     \label{eq:acceptance_critetion}
\end{equation}
This choice implies that if $\Delta E < 0$, the new configuration is always accepted, while if $\Delta E > 0$, it is accepted with probability $\exp(- \Delta E / T )$. This probabilistic acceptance allows the algorithm to escape local minima by occasionally accepting worse configurations. As the temperature gradually decreases according to a suitable cooling schedule \cite{Cohn1999}, this probability diminishes, reducing exploration and encouraging convergence towards a local minimum.

In the \ac{DA}, candidate configurations and their corresponding energy differences are computed in parallel for all variables $x_i \in \vec{x}$ where $i \in \Set{1,2,3,…,n}$. From the set of accepted candidates, one configuration is randomly selected to proceed to the next iteration. This mechanism approximates a rejection-free \cite{kowalsky2022, chen2023, chen2025}, roulette-wheel-style selection process \cite{LIPOWSKI2012}. In a rejection-free MCMC algorithm, every step leads to a state transition. The \ac{DA} further enhances the approximate rejection-free simulated annealing algorithm through three modifications:
\begin{itemize}
	\item \emph{Parallel search}: \ac{DA}v2 performs up to 128 independent simulated annealing runs that are time shared on 16 parallel threads of a single \ac{DAU} (see \cref{appendix:DAv2_runtime_fitting}).
    \ac{DA}v3 uses a single \ac{DAU} together with a global search on multicore CPUs in parallel.
	\item \emph{Dynamic energy offsetting} applies an offset to the energy difference, i.e.,~\\$ \Delta E - \delta E_{\mathrm{off}} $, to increase the acceptance probability in cases where no candidate configurations are accepted, particularly at low temperatures. This helps escape local minima, provided the offset is not excessively large, to avoid losing the progress of the annealing.
	\item \emph{Parallel tempering}, also known as replica exchange: This technique runs multiple independent searches at different temperatures in parallel and exchanges configurations among them to improve energy landscape exploration.
\end{itemize}
In any case, among all assignments evaluated during the algorithm, one with the best objective value is reported as a solution.

The \ac{DAU} is an integer machine, meaning it does not operate natively on floating-point values. When a \ac{QUBO} matrix contains floating-point coefficients, an automatic or manual conversion process must be applied. This involves systematically scaling and rounding the matrix elements to produce integer coefficients that fall within the \ac{DAU}’s supported precision range. Specifically, the \ac{DAU} supports 64-bit signed integers for quadratic terms\footnote{The valid range is $[-2^{63} + 1, 2^{63} - 1]$.} and up to 76-bit signed integers for linear terms. For \ac{DA}v2, the precision is reduced to a 16-bit integer when the problem size exceeds 4,096 variables. It offers both simulated annealing (default) and parallel tempering modes.

The ideas of \ac{DA}v2 are scaled up in \ac{DA}v3, based on a hybrid approach that integrates a global tabu search in a software layer with the \ac{DAU}.
This combination enables the handling of fully connected \ac{QUBO} instances with up to 100,000 variables. This version offers parallel tempering only.

\ac{DA} usage is straightforward: 
the user needs first to convert their problem into a \ac{QUBO} formulation (see the tutorial~\cite{Glover2018}) and then a few lines of code are required to solve it and report the solutions, along with additional details (i.e., energy, time, etc.), see the documentation~\cite{fujitsu_doc, shaglel2025dabenchmark}. 
Fujitsu's \ac{DA} can be easily accessed through a cloud platform or on-premises as a service.

\section{Methodology}
\label{sec:methods}
We compare the performance of \ac{DA}v2 and \ac{DA}v3 against the best-performing classical heuristics selected from a pool of 37 algorithms implemented by \citet{Dunning2018}.  The heuristics are accessible from the corresponding GitHub repository \cite{mqlib} and used in their defaults. 
We will refer to these classical heuristics as \ac{MQLib} heuristics. 
We also use their collection of heterogeneous problem instances, which includes both real-world and random problem instances provided in \cite{mqlib}.

To better understand the strengths of each solver and ensure a reasonable comparison, we split the instance set into 15 groups based on size and density. This grouping allows us to evaluate performance across different problem characteristics and select the top \ac{MQLib} heuristic performers for each category to compare against. The criteria for splitting and selecting the instance set and the selection of the best-of-37 heuristics are described in \cref{subsec:Selection_of_relevant_instances,subsec:best-of-37_heuristic}, respectively. To ensure a representative comparison, we adopt the time-limit assignment methodology from Ref.~\cite{Dunning2018}, which assigns instance-specific time limits reflecting both problem difficulty and hardware improvements over time (\cref{subsec:baseline_timelimit}).

We include complementary comparisons of \ac{DA}v3 with D-Wave’s hybrid quantum-classical solver~\cite{dwave_hybrid_solver_2020}, as well as comparisons of both \ac{DA}v2 and \ac{DA}v3 with the recent quantum-inspired metaheuristic QIS3~\cite{yang2025}.
We include the details of their methodology within the relevant subsections.
While these additional benchmarks are equally important, their setups follow that of the original studies and thus require less detailed methodology discussion.

\subsection{Baseline time limit}
\label{subsec:baseline_timelimit}
\citet{Dunning2018} suggested assigning a specific time limit
to each instance, which reflects its computational difficulty. 
Harder instances are assigned longer time limits based on characteristics such as size and structure, thereby accounting for the varying effort required to reach high-quality solutions. 
By doing so, the benchmark avoids both underestimating the performance of solvers on difficult problems (by giving them too little time) and overestimating performance on easier problems (by giving unnecessarily long runtimes). 
Furthermore, more recent hardware typically requires shorter runtimes on average due to improved processing power compared to older systems. As such, baseline time limits should be periodically updated to reflect current hardware performance.
To make the comparison between the \ac{DA} and the \ac{MQLib} heuristics on the \ac{Max-Cut} instances as fair as possible, we follow Ref.~\cite{Dunning2018} by determining new instance-specific baseline time limits \hyperlink{2a}{(2a)}.

The instance-specific time limit is determined for each instance by the amount of time required to complete the following local search procedure:
\begin{compactenum}
	\item Generate a random configuration for the input instance.
	\item In each updating step, flip the first occurring variable that improves the configuration (i.e., yields a better objective value).
	\item Repeat step 2 until no improving flips can be made.
	\item Repeat steps 1, 2, and 3 for 1500 times.
\end{compactenum}

We executed this algorithm on a Fujitsu PRIMERGY RX2540 M5 server with 2x Intel Xeon Gold 6230 CPU @ 2.10GHz (40 cores total), 12 x 32\,GB = 384\,GB RAM, running RedHat 7.9. 
Due to the algorithm's inherent randomness, the required runtime may vary across different seeds. Therefore, for each instance, we run this algorithm using 5 different seeds and define the time limit as the mean, or 0.001 seconds if the mean falls below this threshold.

\ac{DA}v2 has no time limit stopping criterion and instead terminates when the specified number of runs and iterations are completed. 
Therefore, to set a desired runtime for it, we use a runtime estimate based on the number of variables, runs, and iterations. 
The number of runs is the number of stochastically independent parallel runs on the chip. 
The minimum is 16 runs and the maximum is 128 runs for \ac{DA}v2. The number of iterations is the length of the annealing random walk per run. While the total runtime of \ac{DA}v2 includes several components, it is predominantly determined by the annealing time and the CPU time, which together account for the majority of the runtime. The CPU time refers to the total time spent by the host CPU on tasks such as communicating with the \ac{DAU} and preparing its setup. The annealing time is the actual time spent by the \ac{DA} searching the solution space and finding the optimal or near-optimal solution. Appendix~\ref{appendix:DAv2_runtime_fitting} provides the detailed procedure and fitted runtime models used to estimate the target runtime of \ac{DA}v2.

In contrast, \ac{DA}v3 includes a stopping criterion based on a user-specified time limit. However, as is the case with most solvers, \ac{DA}v3 does not terminate precisely when the specified time limit is reached, as it takes additional time to complete the current search procedure before stopping.
Hence, the runtime tends to overshoot. 
This issue particularly happens for smaller instances, which are typically assigned very short time limits. Notably, \ac{DA}v3 has a minimum runtime of 1 second; if the specified limit falls below this value, the annealing process may not initiate at all. 
To address this issue, we incorporate a buffer time for \ac{DA}v3, aiming to keep the runtime for as many instances to deviate by no more than $10\%$ from the baseline time limit. For instances assigned a time limit below 1 second, we set the minimum to 1 second.
We modify the instance-specific time limit $T_i^{\mathrm{limit}}$ for \ac{DA}v3 such that the runtime stays within the margin of $10\%$: 
\begin{equation}
	{T_i^{\mathrm{limit}}\coloneqq }
	\max\{1, \min\{\lfloor T_i^{\mathrm{baseline}} \rfloor, \lfloor T_i^{\mathrm{baseline}}*1.1 - T_i^\mathrm{offset} \rfloor \} \},
    \label{eq:time_offset}
\end{equation}
where $T_i^{\mathrm{baseline}}$ is the baseline time limit for instance $i$, and $T_i^{\mathrm{offset}}$ is the buffer time, and it is empirically chosen to be 3 seconds. 
Our choice of the offset buffer time is described in Appendix~\ref{appendix:DAv3_offset_selection}. In our analysis, we favor staying below the baseline time limit rather than exceeding it to keep the comparison to other heuristics as fair as possible. 
Appendix~\ref{appendix:runtime_deviation_analysis} presents an analysis of each solver's deviation from the baseline time limit \hyperlink{2b}{(2b)}. Notably, we observed that the \ac{MQLib} heuristics often exceed their specified time limits--an issue not reported in Ref.~\cite{Dunning2018}. Since neither the \ac{DA} nor the heuristics terminate exactly at the assigned time limit, we allow for only very few instances to slightly exceed the predefined safety margin.

As part of the complementary comparisons in our study, we ensure a consistent basis for comparison with D-Wave’s \ac{HS} and QIS3, and we adhere to their fixed time budgets of 20 minutes and 10 seconds per instance, respectively. 
We could not find the reasons why these specific values have been chosen in the respective papers.

\subsection{Selection of graph instances}
\label{subsec:Selection_of_relevant_instances}
Following the benchmark criteria \ref{item:BenchmarkSet} for the benchmark set, we explain and justify below our choice of benchmark set for our main benchmark with \ac{MQLib} heuristics and provide details of the benchmark sets of the additional benchmarks with D-Wave's \ac{HS} and QIS3, as selected by the original works.

\noindent \textbf{DA vs.\ best \ac{MQLib} heuristics}

The full \ac{MQLib} \cite{mqlib} includes instances from the following sources: Gset \cite{ye2003gset}, ORLib \cite{orlib}, SteinLib \cite{steinlib}, 2nd, 7th, and 11th DIMACS \cite{dimacs2,dimacs7,dimacs11}, TSPLib \cite{tsplib}, Biq Mac \cite{biqmac2007}, and more from original solver papers.
Furthermore, it includes random problem instances generated from multiple random generators: Culberson \cite{culberson}, NetworkX \cite{networkx}, and  Rudy \cite{rudy}. At the time of writing this paper, \ac{MQLib} contains a total number of 3506 instances. Exactly 3,396 instances were introduced by the original \ac{MQLib} paper \cite{Dunning2018}, while the remaining instances were later added by external contributors. 
The instances span a wide range in both size $n$ (number of vertices), from as few as 3 to approximately 53,000 variables, and density $d$ (the ratio of actual to possible edges), ranging from extremely sparse graphs with $d \sim 0.0001$ to fully connected graphs with $d = 1$.
\begin{table}[t]
	\centering
	 \resizebox{\textwidth}{!}{
\begin{tabular}{|c|c|c|c|c|c|}
		\hline
		\diagbox[width=\dimexpr \textwidth/6+3\tabcolsep\relax, height=1cm]{Density}{Size} & \makecell{\textbf{x-Small} \\ $ 20 \leq n < 1,024$} & \makecell{\textbf{Small} \\ $ 1,024 \leq n < 2,048$} & \makecell{\textbf{Medium} \\ $2,048 \leq n < 4,096$ } & \makecell{\textbf{Large} \\ $ 4,096 \leq n < 8,192$} & \makecell{\textbf{x-Large} \\ $ 8,192 \leq n $} \\ \hline 
		\makecell{\textbf{Sparse} \\ $d < 0.1$} & 525 & 179 & 279 & 171 & 81 \\ 
		\makecell{\textbf{Balanced} \\ $ 0.1 \leq d < 0.5 $} & 367 & 41 & 78 & 119 & 0 \\ 
		\makecell{\textbf{Dense} \\ $ 0.5 \leq d$}  & 182 & 12 & 50 & 41 & 0 \\ \hline \hline
		\textbf{Total} (2,125) & 1,074 & 232 & 407 & 331 & 81 \\ \hline
		 \textbf{Benchmark} & \ac{DA}v2 & \ac{DA}v2 & \ac{DA}v2 and \ac{DA}v3 & \ac{DA}v2 and \ac{DA}v3 & \ac{DA}v3 \\ \hline
	\end{tabular}}
	\caption{\ac{MQLib} instance counts across categories defined by size and density. The total number of our chosen instances is 2,125. 
	Small and x-Small categories are analyzed in \cref{appendix:small_instances_comparison} for \ac{DA}v2.} 
	\label{table:instances_categories}
\end{table}

From the full instance set, we consider only instances with baseline time limits above 0.25 seconds. This reduces the total number of instances in \ac{MQLib} to 2,125. With this, we filter out particularly easy instances, as well as those for which the solvers are likely to far exceed the assigned time limit due to a particularly short baseline time. 
Most solvers do not check the time limit continuously; instead, they check it after, e.g.~, a certain number of internal iterations, or after completing a certain operation. If the time limit is relatively short, the solver may overshoot it significantly before reaching the next checkpoint.

\begin{figure}[t]
	\centering
	\includegraphics[scale=0.6]{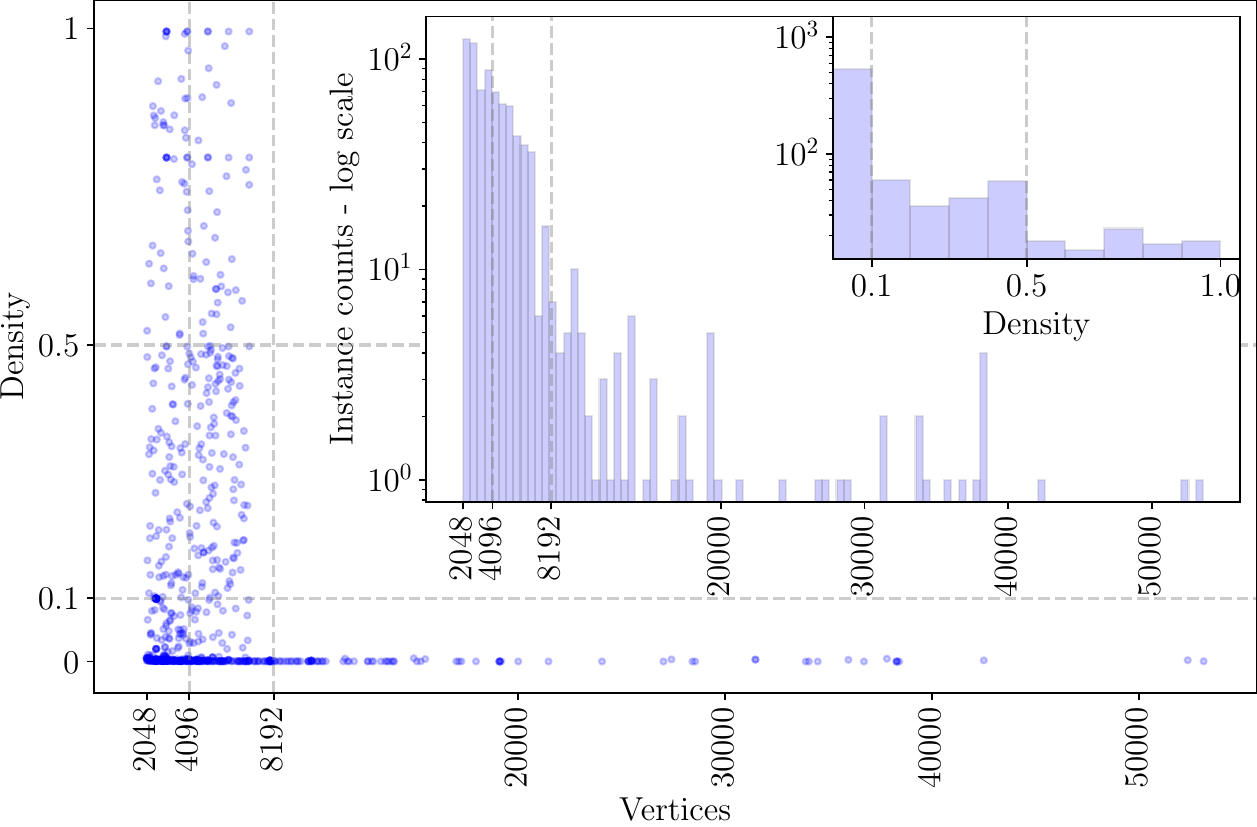}
	\caption{Distribution of instances by number of vertices and density. All instances above $\sim 8,000$ variables are sparse, whereas those with fewer variables exhibit a wider range of densities. Darker blue regions indicate a higher concentration of instances, which can be observed at low density and number of vertices. Inset: Histogram (log scale) showing instance counts by size range; the majority of instances are below $\sim 8,000$ variables. Nested inset: Histogram (log scale) showing instance counts by density range; the majority of instances are sparse.}
	\label{fig:instances_scatter}
\end{figure}

We divide our selected instances into 15 categories based on size and density.
The exact number of instances that fall in each of the size-density category combinations is shown in Table \ref{table:instances_categories}.
In \cref{subsec:DA_vs_mqlib_heuristics}, we compare \ac{DA}v2 on Medium--Large instances (up to what fits the \ac{DAU}) and \ac{DA}v3 on Medium--x-Large instances, across all density categories, against the best-performing \ac{MQLib} heuristic on each category. In these selected categories, the solvers are more likely to adhere to the instance-specific time limits, which helps to ensure the justifiability of the comparison. 
The categories involved in the main analysis are visualized in \cref{fig:instances_scatter} by the regions bounded by dashed lines. Most instances of this selection lie in the Medium- and Large-size categories, with a higher concentration in the Sparse category. The \ac{MQLib} appears to lack x-Large instances that are Balanced or Dense, which explains the empty regions in the middle and upper right of the scatter plot in \cref{fig:instances_scatter}.

\noindent \textbf{DAv3 vs. D-Wave's hybrid solver}

Next, we follow D-Wave's selection of instances from the \ac{MQLib} based on three criteria \cite{dwave_hybrid_solver_2020}. 
First, they consider connected instances only with size between $n=1,000$ and $n=10,000$. Second, all selected instances were reported to have a maximum recommended time limit of 20 minutes. Third, the best objective value of each instance should be achieved by at most two heuristics out of the 37 \ac{MQLib} heuristics. 
These criteria yield a subset of 45 instances ranging from $\sim 2,000$ to 10,000 in size and $\sim 0.0004$ to $\sim 0.97$ in density \cite{dwave_hybrid_solver_2020}. 
We noticed that 39 instances were assigned a time limit below 20 minutes in the original study, and only 6 were assigned a time limit above 20 minutes. 
Moreover, we observed that three instances in this subset
have more than two heuristics achieving the best objective value.  
However, we proceed with D-Wave's selection \cite{dwave_hybrid_solver_2020} and compare \ac{DA}v3 with D-Wave's \ac{HS}.

\noindent \textbf{DA vs. QIS3}

The comparison between \ac{DA}v2, \ac{DA}v3, and the recently introduced quantum-inspired solver QIS3 by \citet{yang2025} is conducted on a subset of 16 instances from the G-set \cite{ye2003gset}, with graph sizes ranging from 800 to 10,000 variables that are very sparse ($\sim$ 0.0004 to 0.06). The G-set comprises a total of 71 instances, but the rationale behind the selection of this particular subset is not discussed.
Notably, two of the selected instances exceed the 8,192-variable capacity of a single \ac{DAU} and therefore cannot be solved using \ac{DA}v2.

\subsection{Choosing the best-performing MQLib heuristics}
\label{subsec:best-of-37_heuristic}
Ref.~\cite{Dunning2018} implements and compares 37 heuristics based on a variety of metaheuristic approaches, including simulated annealing, cross-entropy methods, non-linear optimization, tabu search, global equilibrium search, greedy randomized adaptive search (GRASP), estimation of distribution algorithms, genetic algorithms, and some of their variants. 
In \cref{subsec:baseline_timelimit}, we obtained a new baseline time limit per instance; we therefore aim to compare the \ac{DA} with new runs of \ac{MQLib} heuristics under these time limits. 
For this purpose, we select the best-performing heuristic within each size and density category defined in \cref{subsec:Selection_of_relevant_instances}. 
We define the best heuristic per category as the one that achieves the best objective value for the highest number of instances in the corresponding category, based on the original data by \citet{Dunning2018}. 
These heuristics were then executed on the same server used to establish the baseline time limits and to run the \ac{DA}.
\Cref{table:category_heuristics} lists the selected \ac{MQLib} heuristics for each category, along with the number of instances within that category for which each heuristic found the best solution. BURER2002 is a non-linear optimization method combined with local search \cite{Burer2002RankTwoRH}; PAL2004bMTS2 is an iterated tabu search \cite{Palubeckis2004MultistartTS}; and MERZ1999GLS is a genetic algorithm incorporating crossover and local search \cite{merz1999}.
As shown in \cref{table:category_heuristics}, BURER2002 performs best on Sparse instances while PAL2004bMTS2 is most effective on Balanced and Dense instances. MERZ1999GLS shows the best performance on x-Large Sparse instances. After selecting the best-performing heuristics, we rerun these heuristics in their default settings as provided in \cite{mqlib}
on the relevant instances using the newly assigned time limits.
\begin{table}[t]
	\centering
    \resizebox{\textwidth}{!}{
	\begin{tabular}{|c|c|c|c|c|c|}
		\hline
		\diagbox[width=\dimexpr \textwidth/6+3\tabcolsep\relax, height=0.8cm]{Density}{Size} & \textbf{x-Small} & \textbf{Small} & \textbf{Medium} & \textbf{Large} & \textbf{x-Large} \\ \hline 
		\textbf{Sparse} & \cellcolor{blue!10} 205 (39.0\%)  & \cellcolor{blue!10} 85 (47.5\%)  & \cellcolor{blue!10}  111 (39.8 \%) & \cellcolor{blue!10} 56 (32.7\%) & \cellcolor{green!15} 37 (45.7\%) \\ 
		\textbf{Balanced} & \cellcolor{yellow!20} 183 (49.9\%)  & \cellcolor{yellow!20}  13 (31.7\%) & \cellcolor{yellow!20} 39 (50.0\%) & \cellcolor{yellow!20} 66 (55.5\%) & - \\ 
		\textbf{Dense} & \cellcolor{yellow!20} 119 (65.4\%) & \cellcolor{yellow!20} 6 (50.0\%)  & \cellcolor{yellow!20} 27 (54.0\%)  & \cellcolor{yellow!20} 28 (68.3\%) & - \\ \hline 
	\end{tabular}}
	\caption*{
		\colorbox{blue!10}{\phantom{xx}} BURER2002 \quad
		\colorbox{yellow!20}{\phantom{xx}} PAL2004bMTS2 \quad
		\colorbox{green!15}{\phantom{xx}} MERZ1999GLS
	}
	\caption{Best-performing \ac{MQLib} heuristics across instance categories, color-coded by heuristic. The number in each cell indicates the amount of instances the corresponding heuristic achieved the best objective value for, based on the original data by \citet{Dunning2018}.}
	\label{table:category_heuristics}
\end{table}

 \section{Results}
 \label{sec:results}
In this section, we present the benchmark results for the main study comparing \ac{DA}v2 and \ac{DA}v3 with \ac{MQLib} heuristics in \cref{subsec:DA_vs_mqlib_heuristics}, as well as the complementary comparisons of \ac{DA}v3 with D-Wave's \ac{HS} in \cref{subsec:DAv3_vs_D-Wave}, and of both \ac{DA}v2 and \ac{DA}v3 with QIS3 in \cref{subsec:DA_vs_QIS3}. 
\ac{DA}v3 often exceeds short time limits; therefore, we restrict it to Medium to x-Large instance sizes with relatively longer assigned time limits, which it can to some extent reliably solve within the assigned time, ensuring a meaningful comparison.

For these studies, we executed \ac{DA}v2, \ac{DA}v3, and \ac{MQLib} heuristics on each instance with the specified time limit using five different random seeds, and here we report the highest achieved objective value, along with the minimum runtime among the runs that reached this value \hyperlink{3b}{(3b)}. In addition, whenever applicable, we provide statistical inference using paired run-level significance tests.

To support reproducibility and facilitate further analysis, the GitHub repository accompanying this work \cite{shaglel2025dabenchmark} provides raw data and summary CSV files for each solver considered, listing per-instance minimum, maximum, median, mean, and standard deviation of both the objective values and runtimes across the 5 independent runs.

\ac{DA}v2 was executed in simulated annealing mode, whereas \ac{DA}v3 was executed in its only available mode, parallel tempering. For all instances, minor preprocessing was applied. Specifically, the heaviest variable in an instance (the one with the largest linear coefficient) was fixed, meaning it was pre-assigned to one side of the cut. Additionally, disconnected variables (i.e., $q_{ij} = 0$ for $x_i, x_j \in V$) are automatically excluded by the \ac{DA}, as they do not contribute to the cut. As a result, the resulting \ac{QUBO} problem typically contains $|V|-1$ variables, though in some cases it may involve slightly fewer due to the removal of disconnected variables.

 \subsection{DA vs.\ best MQLib heuristics on Medium--x-Large instances}
 \label{subsec:DA_vs_mqlib_heuristics}
\begin{figure}[t]
 	\centering
 	\includegraphics[width=\linewidth]{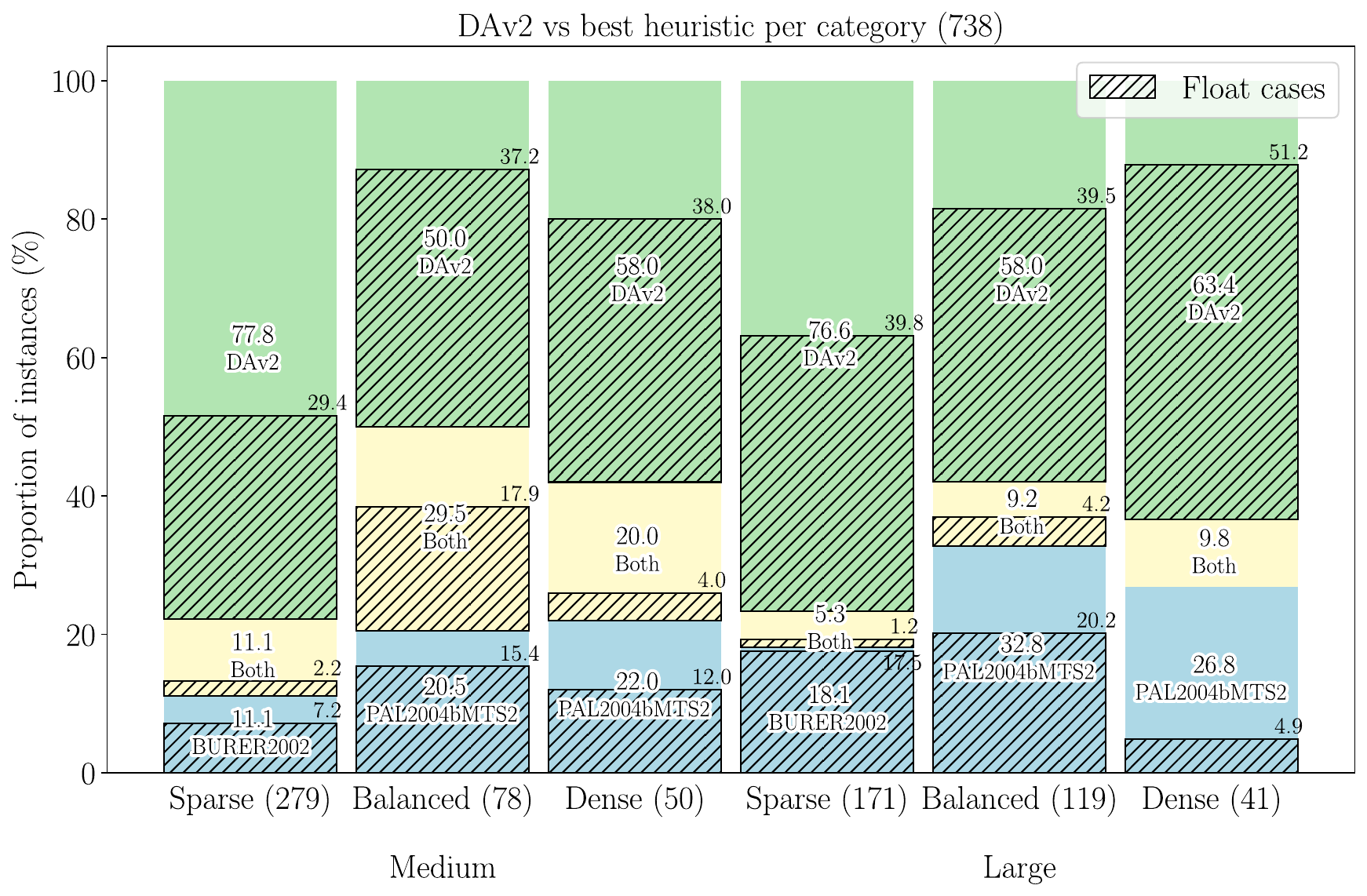}
 	\caption{Comparison of the performance of \ac{DA}v2 with the best-performing category heuristic on Medium and Large instances across all densities. Each bar represents the proportion of instances where \ac{DA}v2 performs better (green), equal (yellow), or worse (blue) than the corresponding heuristic. Hatched bars indicate instances with floating-point cut values.}
 	\label{fig:DAv2_med_large}
\end{figure}

In Fig.~\ref{fig:DAv2_med_large}, we compare \ac{DA}v2 against the best-performing \ac{MQLib} heuristic per category (as selected in \cref{subsec:best-of-37_heuristic}) using a cumulative bar plot. 
Each bar is divided into three segments: proportion of instances where \ac{DA}v2 finds a better cut value than the selected category heuristic (green), proportion of instances where both solvers find the same cut value (yellow), and proportion of instances where the heuristic finds a better cut value than the \ac{DA}v2 (blue). The results show that \ac{DA}v2 yields higher cut values than the selected \ac{MQLib} heuristics on the majority of instances across all categories, with its largest advantage on Sparse instances--achieving greater cut values than BURER2002 on over 75\% of the instances.
\begin{figure}[H]
 	\centering
 	\includegraphics[width=\linewidth]{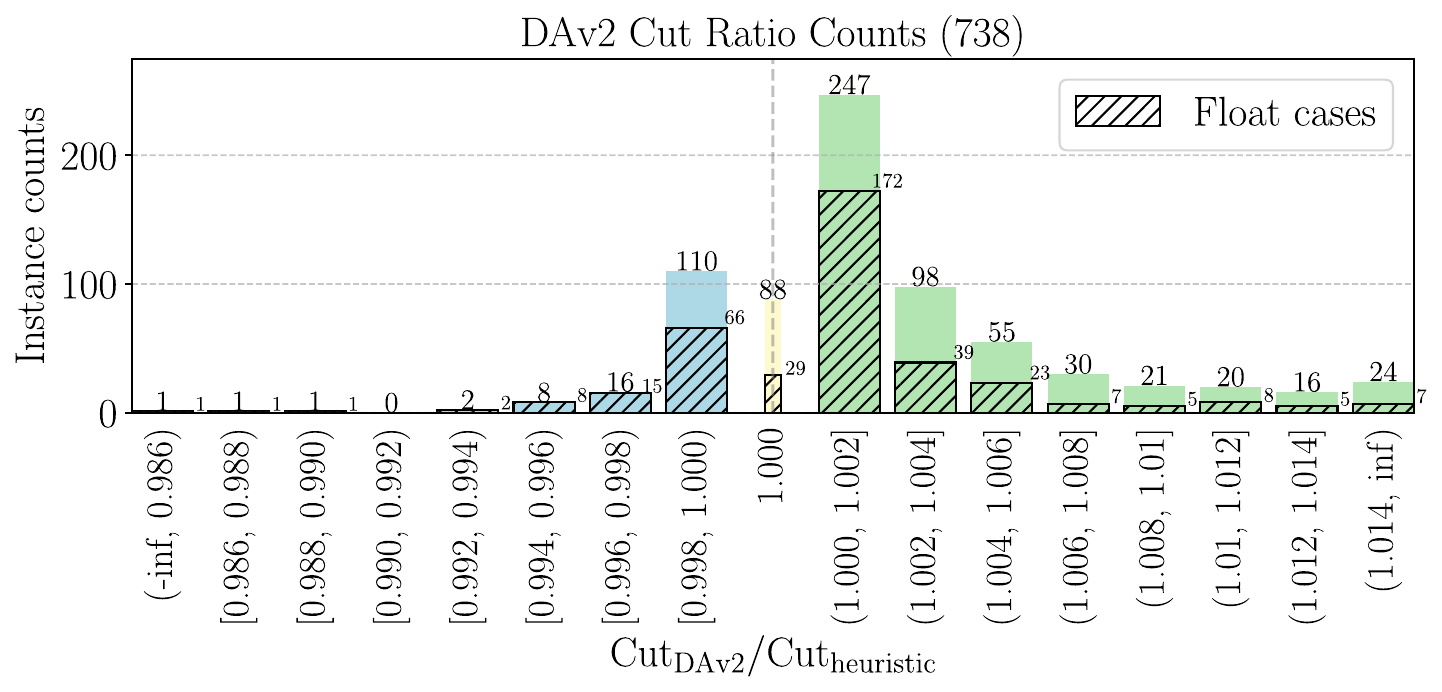}
 	\caption{Distribution of Medium and Large instances over cut ratio ranges, where the cut ratio is defined as the \ac{DA}v2 cut divided by that of the best-performing category heuristic. The `tie' bar at 1.0 (yellow) shows the count of instances when both solvers found an equal cut value. The dashed line at 1.0 separates `win' ranges (green bars) from `loss' ranges (blue bars). Hatched bars indicate instances with floating-point cut values.}
 	\label{fig:DAv2_acc_counts}
\end{figure}

\newpage
Fig.~\ref{fig:DAv2_acc_counts} presents a bar plot of instance counts across cut ratio ranges, defined as the cut value achieved by \ac{DA}v2 to that achieved by the corresponding heuristic. The plot shows a clear skew to values greater than 1.0 (i.e., higher green bars), highlighting an overall better result of \ac{DA}v2 than the best-performing category heuristics. Nearly all instances with cut ratio below 0.998 are float-type instances.
The \ac{DA} natively operates on integer-valued \ac{QUBO} coefficients. Instances with floating-point values require scaling and rounding before processing. Therefore, when the range between the smallest and largest \ac{QUBO} coefficient values $q_{ij}$ of an instance is too wide, scaling the coefficients up by $2^{\alpha}/\max(|q_{ij}|)$ with $\alpha = 63$ or $\alpha=15$ for \ac{DA}v2 (depending on the size of the \ac{QUBO} matrix) and $\alpha = 48$ for \ac{DA}v3 may still result in the smallest positive coefficients having values below 1. These are then rounded down to the nearest integer, and in this case, to zero, which results in the loss of important problem information and distorts the original \ac{QUBO} problem.

In total, based on best-of-5-runs, \ac{DA}v2 achieves better cuts in 511 and equal in 88 out of 738 instances, resulting in a `win' rate of $\sim$ 69.24\%, a `tie' rate of $\sim 11.92\%$, and a `loss' rate of $\sim 18.83\%$. 
To assess statistical significance we consider paired (by seed) run-level comparisons. \ac{DA}v2 achieves better objective values than the corresponding category heuristics in a majority ($\geq 3$) of five paired runs on 542 of 738 instances ($\sim$ 73.44\%), which is highly significant according to a two-sided paired sign (binomial) test ($p = 2.35 \times 10^{-38}$). 
Runs in which \ac{DA}v2 and the corresponding category heuristics achieve identical objective values are conservatively counted in favor of the latter. This strict criterion constitutes a worst-case assumption for \ac{DA}v2's performance and reduces the risk of overstating its advantage. The significantly low $p$-value shows that the observed dominance of \ac{DA}v2 cannot plausibly be attributed to random variation.

\begin{figure}[t]
	\centering
	\includegraphics[width=\linewidth]{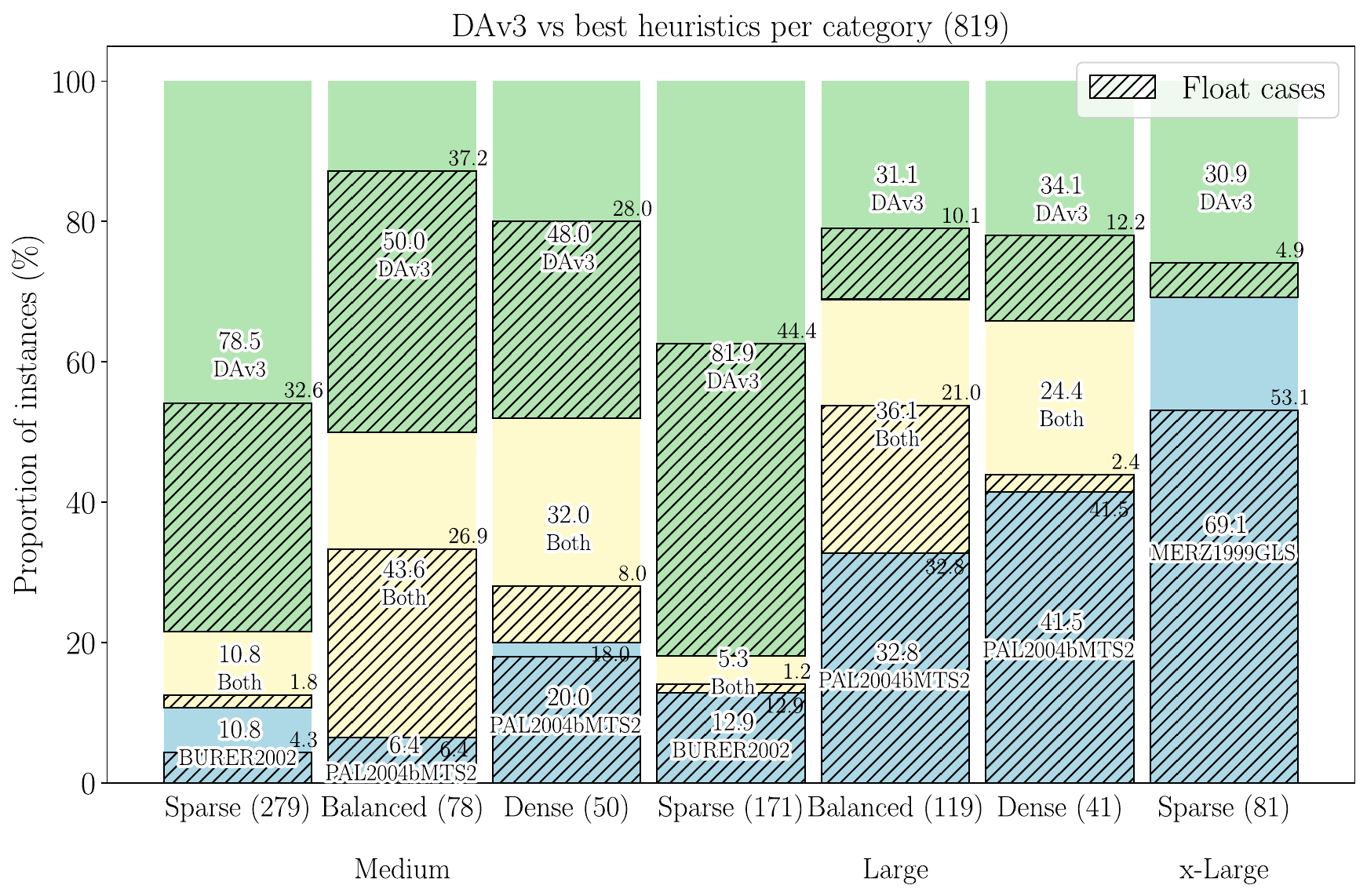}
	\caption{Comparison of the performance of \ac{DA}v3 against the best-performing category heuristic on Medium, Large, and x-Large instances across all available densities. Each bar represents the proportion of instances where \ac{DA}v3 performs better (green), equal (yellow), or worse (blue) than the corresponding heuristic. Hatched bars indicate instances with floating-point cut values.}
	\label{fig:DAv3_med_xlarge}
\end{figure}

We now turn to \ac{DA}v3 and evaluate its performance on the Medium to x-Large instance categories using the same methodology. As shown in Fig.~\ref{fig:DAv3_med_xlarge}, \ac{DA}v3 clearly achieves higher quality results than BURER2002 on Sparse instances, while PAL2004bMTS2 surpasses \ac{DA}v3 on Large instances that are Balanced and Dense. 
The majority of `wins' by the selected heuristics correspond to float-type instances, as indicated by the hatched bars--possibly due to the effect of rounding errors on the \ac{DA}'s cut values. In fact, the rounding error is upper bounded as $|x^\top Qx - x^\top\tilde{Q}x| \leq 2^{-\alpha}n^2\max(|q_{ij}|)$, where $\alpha = 48$ for \ac{DA}v3 and $\tilde{Q}$ is the scaled and rounded \ac{QUBO} matrix with coefficients $\tilde{q}_{ij} = \frac{\lfloor q_{ij}*2^{\alpha}/\max(|q_{ij}|) \rfloor }{2^{\alpha} / \max{(|q_{ij}|)}}$. This error bound explains the sensitivity of the \ac{DA} to rounding errors with increasing instance sizes. Additionally, MERZ1999GLS shows a notable advantage over \ac{DA}v3 on the x-Large Sparse category.

\begin{figure}[H]
 	\centering
 	\includegraphics[width=\linewidth]{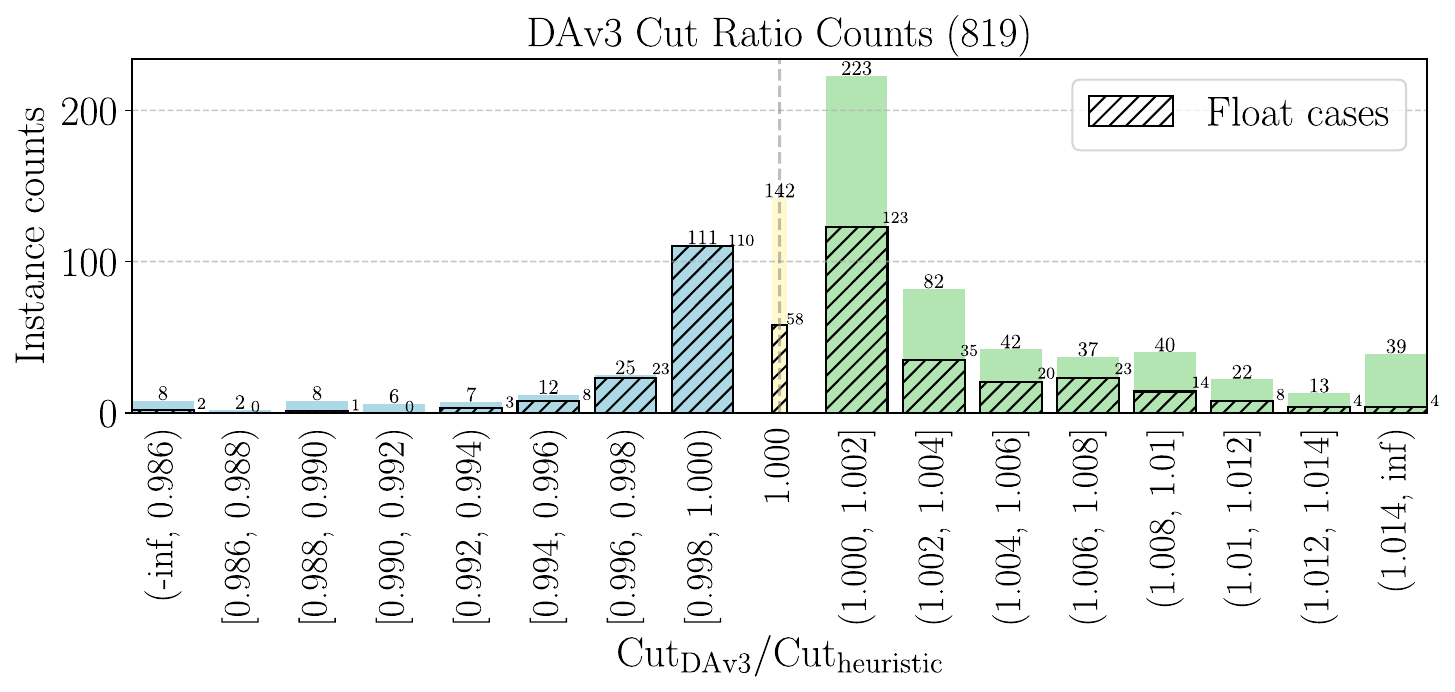}
 	\caption{Distribution of Medium, Large, and x-Large instances over cut ratio ranges, where the cut ratio is defined as the \ac{DA}v3 cut divided by that of the best-performing category heuristic. The `tie' bar at 1.0 (yellow) shows the count of instances when both solvers found an equal cut value. The dashed line at 1.0 separates `win' ranges (green bars) from `loss' ranges (blue bars). Hatched bars indicate instances with floating-point cut values.}
	\label{fig:DAv3_acc_counts}
\end{figure}

Compared to \ac{DA}v2, both versions perform best on Medium and Large Sparse instances, with \ac{DA}v3 showing a slight improvement. However, \ac{DA}v2 achieves 26.9\% and 29.3\% improved results on Large Balanced and Large Dense instances, respectively.
Similar to \ac{DA}v2, the cut ratio bar plot for \ac{DA}v3 in Fig.~\ref{fig:DAv3_acc_counts} shows a right-skewed distribution, indicating overall improved results compared to the best-performing category heuristics. However, unlike \ac{DA}v2, most \ac{DA}v3's `losses' on float cases fall within $[0.998, 1)$ cut ratio range, suggesting a greater sensitivity to rounding errors. 

In total, \ac{DA}v3 achieves better cuts in 498 and equal in 142 out of 819 instances, resulting in a `win' rate of $\sim 60.81\%$, `tie' rate of $\sim 17.34\%$, and a `loss' rate of $\sim 21.85\%$. Using the same paired run-level comparison procedure described above, \ac{DA}v3 achieves better objective values in a majority ($\geq 3$) of five paired runs on 517 of 819 instances ($\sim$ 63.13\%), which is again highly significant according to a two-sided paired sign (binomial) test ($p \sim 5.45 \times 10^{-14}$). 
\begin{figure}[t]
 	\centering
	\begin{subfigure}{\textwidth}
		\includegraphics[width=\linewidth]{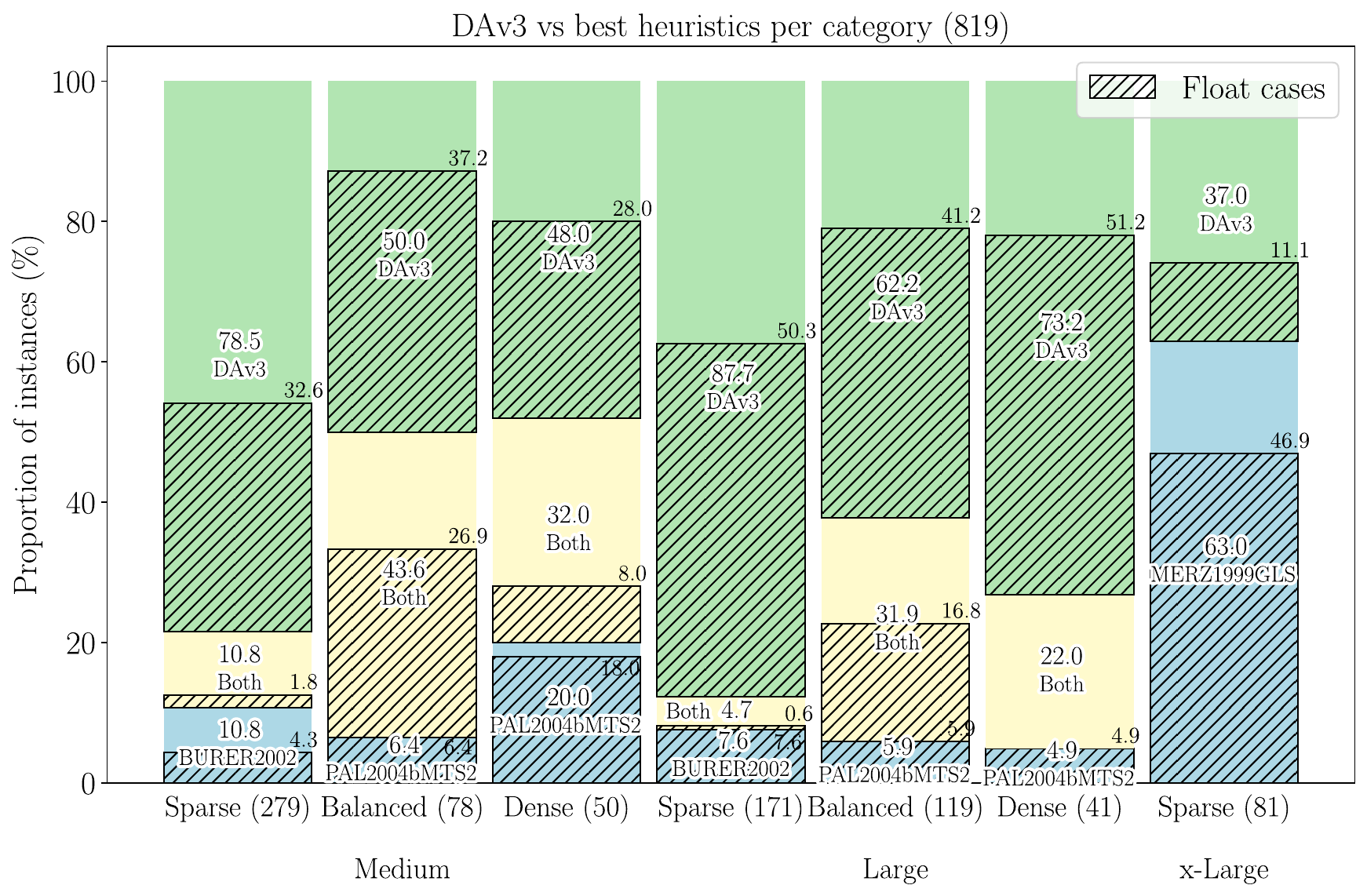}
	\end{subfigure}
	
	\begin{subfigure}{\textwidth}
		\includegraphics[width=\linewidth]{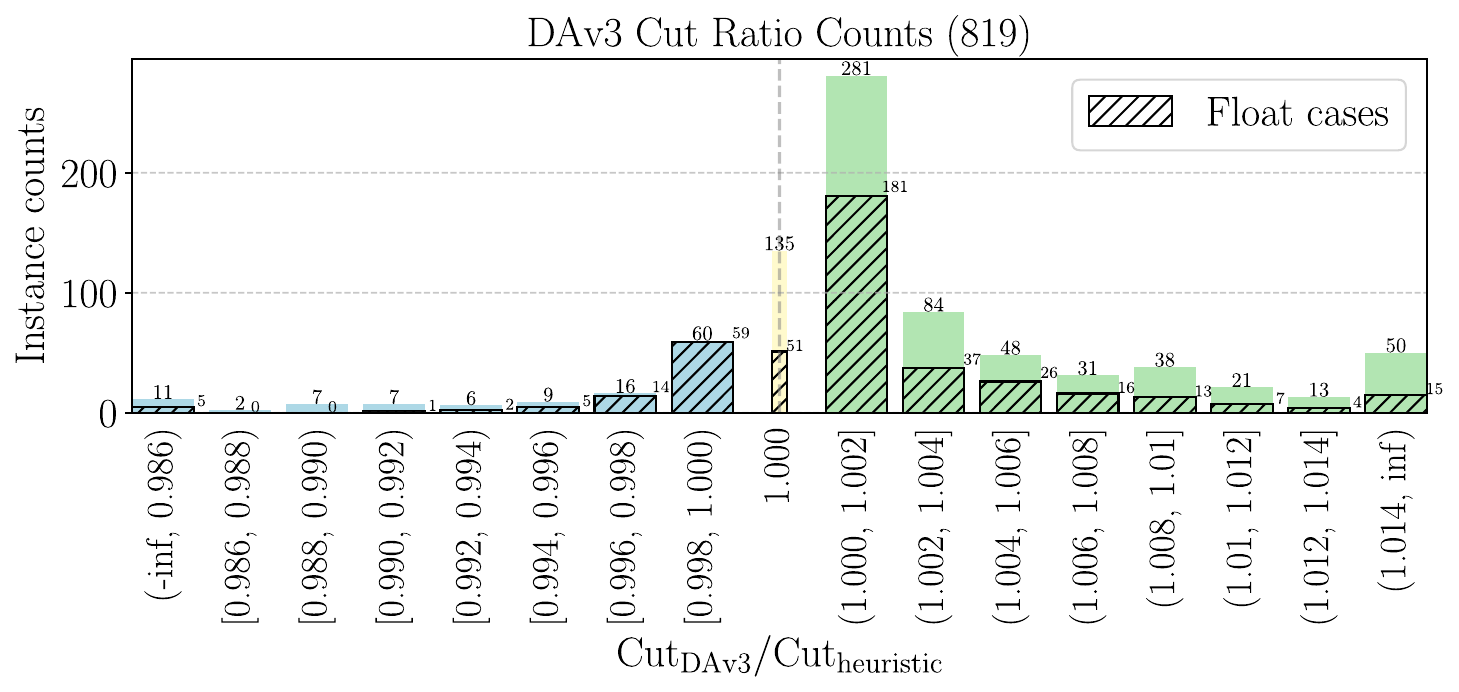}
	\end{subfigure}
    
 	\caption{Improved results of \ac{DA}v3 by the manual scaling described in \cref{alg:scaling-rule} instead of the internal automatic scaling.}
 	\label{fig:DAv3_improved}
 \end{figure}
 
One advantage of partitioning the instance set by size is that it revealed a significant, initially unexplained degradation in the solution quality of \ac{DA}v3 on the instances in the Large category. Upon investigation, we found that the internal automatic scaling of \ac{DA}v3 is not efficient. To address this, we applied manual scaling. Our observations indicate that with 48-bit precision and instance sizes larger than 4,096 vertices, the solver partitions the problem into 4,096-variable chunks, which leads to degraded performance. By manually setting the precision to 64 bits\footnote{Later scaled down internally by \ac{DA}v3 to 48-bit precision.} for $n \leq 4,096$ and 16 bits for $n > 4,096$ (mimicking the behavior of \ac{DA}v2), instances larger than 4,096 vertices (still smaller than 8,192) can be embedded as a single chunk and solved in full, resulting in better performance as shown in \cref{fig:DAv3_improved} above. 

Thus, we provide an algorithmic rule (\cref{alg:scaling-rule})
for coefficient scaling and a recommendation on when to trust the automatic scaling and when to choose the manual scaling.

The results confirm a substantial improvement for the Large category across all densities: the proportion of instances where \ac{DA}v3 found a better cut increased by 5.8\% for Sparse instances, 31.1\% for Balanced instances, and 39.1\% for Dense instances. A slight gain (6.1\%) was also observed for x-Large Sparse instances, though larger improvements are not expected due to the unavoidable partitioning of the problem into 8,192-variable chunks. Overall, with this adjustment, \ac{DA}v3 achieved better cuts in 566 and equal in 135 out of 819, resulting in a `win' rate of $\sim 69.11\%$, `tie' rate of $\sim 16.48\%$, and a `loss' rate of $\sim 14.41\%$.
Using the same paired run-level comparison procedure as before, \ac{DA}v3 achieves better objective values in a majority ($\geq 3$) of five paired runs on 582 of 819 instances ($\sim$ 71.06\%), which is again highly significant according to a two-sided paired sign (binomial) test ($p \sim 2.81 \times 10^{-34}$).
\begin{algorithm}[t]
  \caption{Selection rule for \ac{DA}v3 scaling setting}
  \label{alg:scaling-rule}
  \DontPrintSemicolon
  \KwIn{Instance size $n$ and coefficient-type (int-only / float)}
  \KwOut{Scaling setting for \ac{DA}v3}

  \uIf{$ n \le 4,096$}{
    \uIf{int-only}{
      Use integer representation without scaling.\;
      \tcp*[f]{Scaling not needed}\;
      \tcp*[f]{All coefficients fit into int64}\;
    }
    \uElseIf{float}{
      Scale coefficients up so that they fit into int64 \texttt{OR} use \ac{DA}v3 automatic scaling.\;
      \tcp*[f]{Automatic scaling can be trusted in this regime}\;
    }
  }
  \ElseIf{$n > 4,096$}{
    \uIf{int-only}{
      \uIf{max coefficient fits into int16}{
        Use integer representation without scaling.\;
          \tcp*[f]{Scaling not needed}\;
          \tcp*[f]{All coefficients fit into int16}\;
      }
      \Else{
        Scale coefficients down so that they fit into int16.\;
        \tcp*[f]{Automatic scaling can be outperformed}\;
      }
    }
    \uElseIf{float}{
      Scale (up or down) coefficients so that they fit into int16.\;
      \tcp*[f]{Automatic scaling can be outperformed
      }\;
    }
  }

  Run \ac{DA}v3 with the selected scaling setting.\;
\end{algorithm}

\cref{table:scaling_improve} presents the changes in the cut ratios (i.e.~ $\mathrm{Cut}_\mathrm{DAv3}/\mathrm{Cut}_\mathrm{heuristic}$) in greater detail before and after applying our manual scaling (\cref{alg:scaling-rule})
in place of the automatic scaling used by \ac{DA}v3. 
The diagonal entries of the table correspond to instance counts that remain within the same cut ratio range. Counts below the diagonal indicate improvements in the cut ratio, whereas counts above the diagonal indicate degradations. As shown in the table, the majority of changed instance counts lie below the diagonal, demonstrating improved performance under manual scaling. This finding highlights a trade-off between integer precision and problem decomposition, with greater gains achieved by solving the problem as a whole at lower integer precision.

Nevertheless, for consistency, we continue the following comparisons with D-Wave's \ac{HS} and QIS3 using the default automatic scaling of \ac{DA}v3.

\begin{table}[t]
\centering
\footnotesize
\resizebox{\textwidth}{!}{
\begin{tabular}{|l| l*{16}{c} | c 
  |}
\hline
\diagbox{to}{from}
& \rotatebox{90}{(-inf, 0.986)}
& \rotatebox{90}{[0.986, 0.988)  }
& \rotatebox{90}{[0.988, 0.990)  }
& \rotatebox{90}{[0.990, 0.992)}
& \rotatebox{90}{[0.992, 0.994)}
& \rotatebox{90}{[0.994, 0.996)}
& \rotatebox{90}{[0.996, 0.998)}
& \rotatebox{90}{[0.998, 1.000)}
& \rotatebox{90}{{1.000}}
& \rotatebox{90}{(1.000, 1.002]}
& \rotatebox{90}{(1.002, 1.004]}
& \rotatebox{90}{(1.004, 1.006]}
& \rotatebox{90}{(1.006, 1.008]}
& \rotatebox{90}{(1.008, 1.01]}
& \rotatebox{90}{(1.01, 1.012]}
& \rotatebox{90}{(1.012, 1.014]}
& \rotatebox{90}{(1.014, inf)}
& 
\rotatebox{90}{Total (after)} \\
\hline

(-inf, 0.986) & 8 &  &  &  &  &  &  & 2 & 1 &  &  &  &  &  &  &  &  & 11 \\
{}[0.986, 0.988) &  & 2 &  &  &  &  &  &  &  &  &  &  &  &  &  &  &  & 2 \\
{}[0.988, 0.990) &  &  & 7 &  &  &  &  &  &  &  &  &  &  &  &  &  &  & 7 \\
{}[0.990, 0.992) &  &  & 1 & 6 &  &  &  &  &  &  &  &  &  &  &  &  &  & 7 \\
{}[0.992, 0.994) &  &  &  &  & 5 &  &  & 1 &  &  &  &  &  &  &  &  &  & 6 \\
{}[0.994, 0.996) &  &  &  &  & 2 & 5 & 2 &  &  &  &  &  &  &  &  &  &  & 9 \\
{}[0.996, 0.998) &  &  &  &  &  & 4 & 12 &  &  &  &  &  &  &  &  &  &  & 16 \\
{}[0.998, 1.000) &  &  &  &  &  &  & 4 & 48 & 6 & 2 &  &  &  &  &  &  &  & 60 \\
{1.000} &  &  &  &  &  &  &  &  & 135 &  &  &  &  &  &  &  &  & 135 \\
(1.000, 1.002] &  &  &  &  &  &  & 3 & 58 &  & 220 &  &  &  &  &  &  &  & 281 \\
(1.002, 1.004] &  &  &  &  &  & 1 & 3 & 1 &  & 1 & 78 &  &  &  &  &  &  & 84 \\
(1.004, 1.006] &  &  &  &  &  & 2 & 1 & 1 &  &  & 4 & 39 & 1 &  &  &  &  & 48 \\
(1.006, 1.008] &  &  &  &  &  &  &  &  &  &  &  & 2 & 28 & 1 &  &  &  & 31 \\
(1.008, 1.01] &  &  &  &  &  &  &  &  &  &  &  &  & 5 & 33 &  &  &  & 38 \\
(1.01, 1.012] &  &  &  &  &  &  &  &  &  &  &  &  &  & 5 & 16 &  &  & 21 \\
(1.012, 1.014] &  &  &  &  &  &  &  &  &  &  &  &  & 1 &  & 2 & 10 &  & 13 \\
(1.014, inf) &  &  &  &  &  &  &  &  &  &  &  & 1 & 2 & 1 & 4 & 3 & 39 & 50 \\
\hline

Total (before) & 8 & 2 & 8 & 6 & 7 & 12 & 25 & 111 & 142 & 223 & 82 & 42 & 37 & 40 & 22 & 13 & 39 & 819 \\
\hline

\end{tabular}}

\caption{Changes in cut-ratio distributions, shown as counts per cut-ratio range, before and after replacing the automatic scaling of \ac{DA}v3 with the manual scaling rule in \cref{alg:scaling-rule}. Diagonal entries indicate unchanged cut ratios (remaining in the same range), entries below (above) the diagonal indicate improvements (degradations).}

\label{table:scaling_improve}
\end{table}

 \subsection{DAv3 vs. D-Wave's hybrid solver}
 \label{subsec:DAv3_vs_D-Wave}
\begin{figure}[t]
 	\centering
	\begin{subfigure}{0.7\textwidth}
		\includegraphics[width=\textwidth]{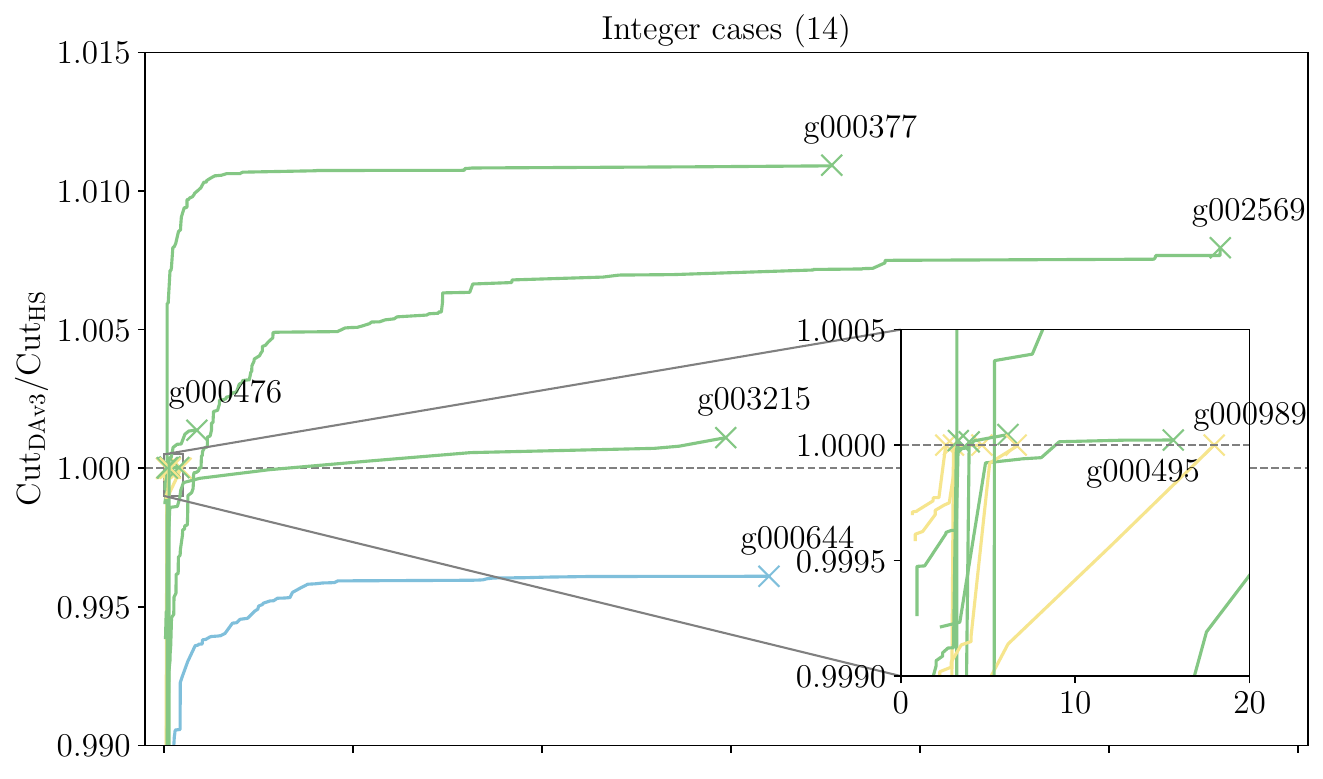}
		\label{fig:int_progress}
	\end{subfigure}
	\hfill
	\begin{subfigure}{0.28\textwidth}
		\includegraphics[width=\textwidth]{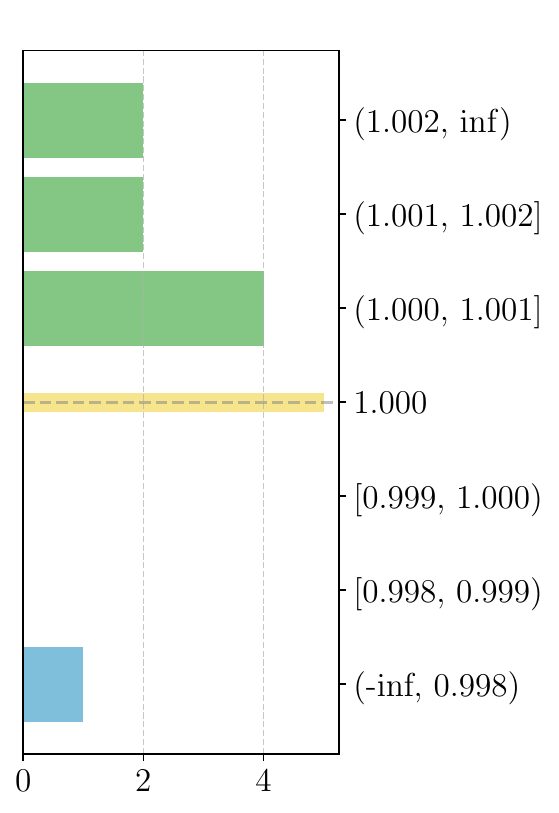}
		\label{fig:int_accuracy_ratio}
	\end{subfigure}
	
	\begin{subfigure}{0.709\textwidth}
		\includegraphics[width=\textwidth]{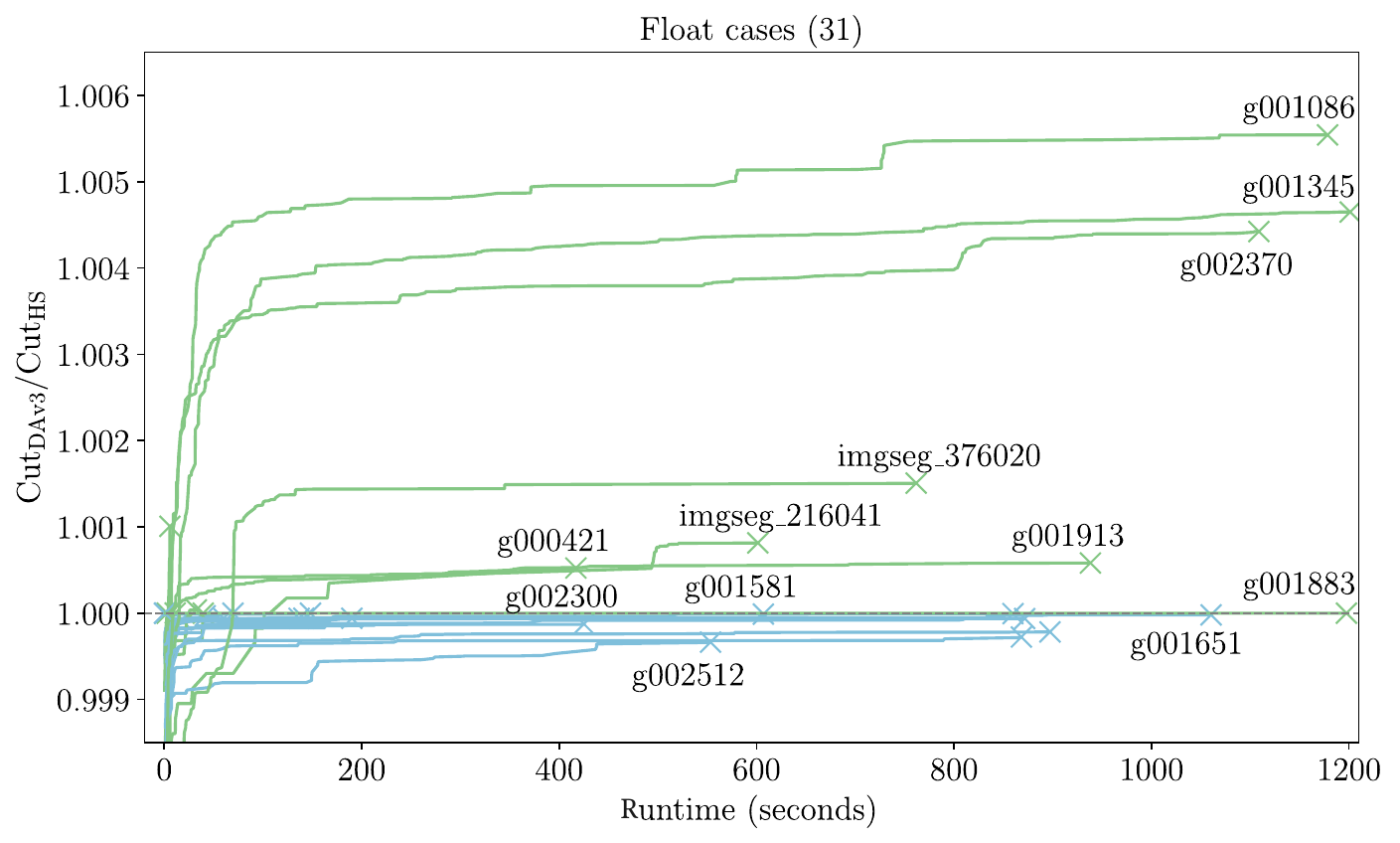}
		\label{fig:flt_progress}
	\end{subfigure}
	\hfill
	\begin{subfigure}{0.28\textwidth}
		\includegraphics[width=\textwidth]{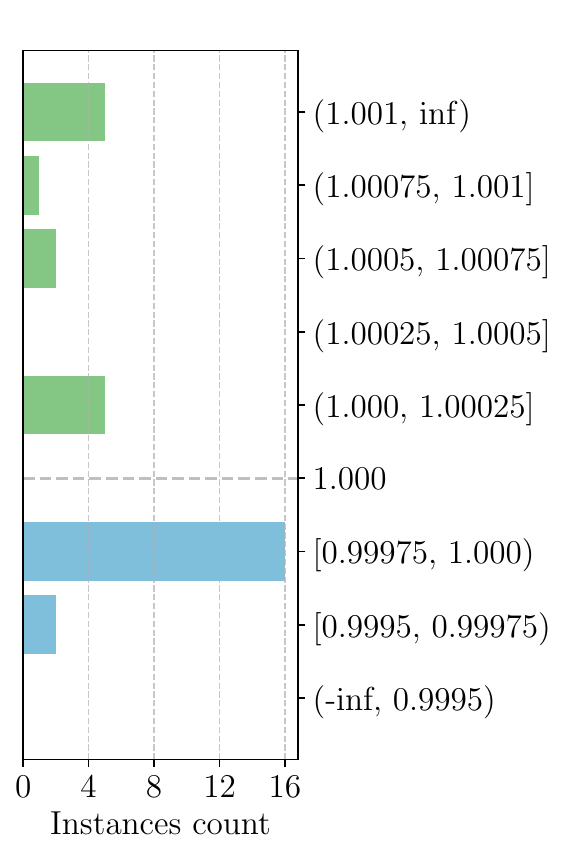}
		\label{fig:flt_accuracy_ratio}
	\end{subfigure}

 	\caption{Left: Progress of \ac{DA}v3 cut ratio relative to D-Wave's \ac{HS} over the 20-minute time limit, separated by integer (top) and float (bottom) instance sets. Each line corresponds to an individual instance, with crosses marking the time when the best cut was found. Colors indicate the performance of \ac{DA}v3 relative to D-Wave's \ac{HS}: green for better, yellow for equal, and blue for worse. Right: Bar plot summarizing final cut ratio distribution.}
 	\label{fig:int_flt_progress_DAv3}
 \end{figure}

 We benchmark \ac{DA}v3 against D-Wave's \ac{HS} \cite{dwave_hybrid_solver_2020} using their specified instance selection criteria, as also detailed in \cref{subsec:Selection_of_relevant_instances}. The test set is further divided into integer (14 instances) and float (31 instances) cases. Figure~\ref{fig:int_flt_progress_DAv3} shows the progress of \ac{DA}v3 solution quality relative to D-Wave's \ac{HS}'s solution, over a 20-minute time limit (left), alongside a summary cut ratio bar plot (right). The upper plots correspond to integer cases, and the lower plots to float cases. Each line corresponds to a given instance (best-of-5 runs), with crosses marking the time the final reported solution was found by \ac{DA}v3.

 For integer cases, \ac{DA}v3 outperforms D-Wave \ac{HS} with 8 `wins', 5 `ties', and 1 `loss'. The progress inset shows that for most instances, \ac{DA}v3 finds an equal or better solution within the first few seconds. Under the tie-handling scheme in which paired runs yielding identical objective values contribute to 0.5 to each solver, an instance is counted as a win for \ac{DA}v3 if its aggregated run-level score exceeds 2.5. According to this criterion, \ac{DA}v3 outperforms D-Wave's \ac{HS} on 9 of 14 integer instances ($\sim$ 64.29 \%). However, this difference is not statistically significant according to a two-sided paired sign (binomial) test ($p \sim 0.424$), indicating that the observed `win' imbalance is compatible with random variation.
 
 For float instances, \ac{DA}v3 underperforms D-Wave's \ac{HS} on 18 instances and outperforms on 13 instances. Most `losses' cluster near a cut ratio range of 0.99975 to 1. This issue is expected given the integer nature of the machine, i.e., encoding the coefficient of a binary quadratic polynomial as an integer. Notably, for many float instances, \ac{DA}v3 finds better solutions in less than 20 minutes, with 5 instances in just a few seconds. \ac{DA}v3 achieves better instance-level results on 10 of 31 float instances ($\sim$ 32.26 \%). The corresponding two-sided paired sign (binomial) test does not indicate a statistically significant difference ($p \sim 0.071$), suggesting that performance differences remain inconclusive. 
 Applying the manual scaling discussed in the previous section improves the results of \ac{DA}v3 on the float-instances, reducing the number of `losses' from 18 to 13.
 
 The rapid convergence of \ac{DA}v3 \hyperlink{3a}{(3a)} to its best solution, often matching or surpassing those of D-Wave's \ac{HS}, highlights the practical potential of for runtime-sensitive scenarios.

\subsection{\texorpdfstring{\ac{DA}}{DA} vs. QIS3}
\label{subsec:DA_vs_QIS3}
During the preparation of this manuscript, a new metaheuristic \ac{QUBO} solver, QIS3, was introduced by \citet{yang2025}. QIS3 combines branch-and-bound pruning with gradient-descent refinement for intensified local exploration, and a quantum annealing-inspired bounding technique to accelerate the pruning. It also incorporates adaptive strategies to optimize performance across diverse problem instances. In their work, the authors benchmark QIS3 on a subset of 16 instances from the G-set collection \cite{ye2003gset}, against eight state-of-the-art solvers: their previous version QIS2, a genetic algorithm, the coherent Ising machine, simulated annealing, parallel tempering, simulated bifurcation, D-Wave's simulated annealing (Neal), and Gurobi. All algorithms were evaluated under 10-second runtime constraints.
Therefore, we configure the parameters of \ac{DA}v2 as described in \cref{appendix:DAv2_runtime_fitting} to achieve a runtime of approximately 10 seconds. For \ac{DA}v3, we run the solver with several time offsets ranging from 0 to 9 seconds and select the offset that yields a runtime closest to 10 seconds. 
In their original study, QIS3 reported the highest cut value on 15 out of 16 instances (with the exception of instance G58).

 \begin{table}[t]
	\resizebox{\textwidth}{!}{
		\begin{tabular}{lcc|c|cc|ccc}
			& \multicolumn{2}{c}{} & \multicolumn{1}{l}{QIS3} & \multicolumn{2}{l}{DAv2} & \multicolumn{3}{l}{DAv3} \\ 
			\cline{4-4} \cline{5-5} \cline{6-9} \\
			name & $n$ & $d$ & cut  & cut & time (sec) & cut & time (sec) & limit (sec)  \\ \hline
			G11 & 800 & 0.005 &  \bfseries 564  &  \bfseries 564 & 9.866  &  \bfseries 564 & 10.983 & 10 \\
			G32 & 2000 & 0.002 & 1404  &  \bfseries 1410 & 9.91  &  \bfseries 1410 & 9.652 & 8 \\
			G48 & 3000 & 0.0013 &  \bfseries 6000  &  \bfseries 6000 & 9.838 &  \bfseries 6000 & 8.795 & 8 \\
			G57 & 5000 & 0.0008 & 3466  & 3470 & 9.959  &  \bfseries 3482 & 8.737 & 6 \\
			G62 & 7000 & 0.0006 & 4828  & 4838& 9.961  &  \bfseries 4846 & 10.759 & 7 \\
			G65 & 8000 & 0.0005 & 5502  & 5460 & 10.139  &  \bfseries 5534 & 12.166 & 8 \\
			G66 & 9000 & 0.0004 &  \bfseries 6288  & - & -  & 6180 & 8.772 & 8 \\
			G72 & 10000 & 0.0004 &  \bfseries 6916 & - & - & 6728 & 9.029 & 4 \\
			G14 & 800 & 0.0147 & 3060  &  \bfseries 3064 & 9.866  &  \bfseries 3064 & 10.978 & 10 \\
			G51 & 1000 & 0.0118 & 3846  &  \bfseries 3848 & 9.869  &  \bfseries 3848 & 11.034 & 9 \\
			G35 & 2000 & 0.0059 & 7673  &  \bfseries 7686 & 9.903  &  \bfseries 7686 & 9.635 & 8 \\
			G58 & 5000 & 0.0024 & 19216  & 19262 & 9.956  &  \bfseries 19263 & 8.893 & 8 \\
			G63 & 7000 & 0.0017 & 26949  &  \bfseries 27012 & 9.942  & 27003 & 10.873 & 8 \\
			G1 & 800 & 0.06 &  \bfseries 11624  &  \bfseries 11624 & 9.864  &  \bfseries 11624 & 10.995 & 10 \\
			G43 & 1000 & 0.02 &  \bfseries 6660  &  \bfseries 6660 & 9.871  &  \bfseries 6660 & 11.041 & 9 \\
			G22 & 2000 & 0.01 & 13358  &  \bfseries 13359 & 9.901  &  \bfseries 13359 & 9.725 & 9 \\ \hline
	\end{tabular}}
	\caption{Comparison of \ac{DA}v2, \ac{DA}v3, and QIS3 on G-set instances \cite{ye2003gset}. The results for QIS3 are taken from \citet{yang2025}, while the \ac{DA} solvers were benchmarked under similar 10-second runtime constraints. The actual runtime for \ac{DA}v2 and \ac{DA}v3 is reported for each instance. G66 and G72 exceed the 8,192-variable capacity of \ac{DA}v2 and thus yield no results for that version.}
	\label{table:QIS3vsDA}
\end{table}

Results of \ac{DA}v2, \ac{DA}v3, and QIS3 (as reported in Ref.~\cite{yang2025}) are presented in \cref{table:QIS3vsDA}. For the \ac{DA}, both the time limit and actual runtimes are reported \hyperlink{2b}{(2b)}, while only the time limit is available for QIS3. Notably, one or both versions of the \ac{DA} find better or matching results than QIS3--and thus than all seven solvers in Ref.~\cite{yang2025}-- on 14 out of the 16 instances, including G58. The only exceptions are instances G66 and G72, where QIS3 remains the top performer, and G65, where only \ac{DA}v3 found a better cut than QIS3. The drop in performance for \ac{DA}v3 on G66 and G72 is attributed to the required partitioning when the problem size exceeds the 8,192-variable capacity of a single \ac{DAU}.

Comparing both \ac{DA} versions across the 14 instances where results are available for both, \ac{DA}v3 yields better results more frequently than \ac{DA}v2. \ac{DA}v2 finds a larger cut than \ac{DA}v3 only on G63, while it fails against both \ac{DA}v3 and QIS3 on G65.

\begin{figure}[t]
	\centering
	\includegraphics[width=\linewidth]{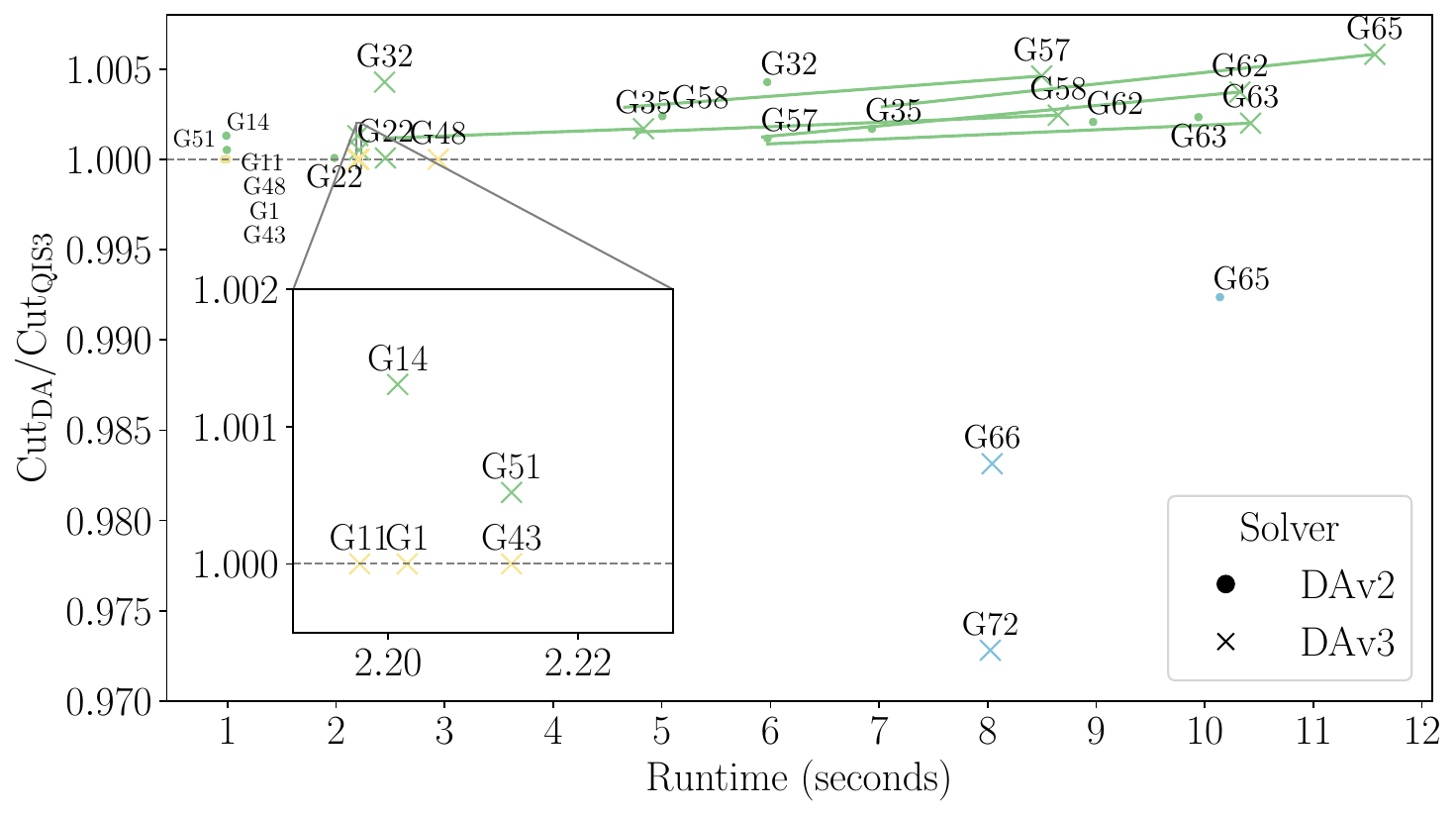}
	\caption{Progress of \ac{DA}v3 solution quality relative to QIS3 over the 10 seconds time limit. Each line corresponds to an individual instance, with crosses marking the time when the best solution was found. Each dot corresponds to the cut quality of \ac{DA}v2 on a given instance, relative to QIS3, measured at the earliest time the best cut was found across independent runs ranging from 1 to 10 seconds. Colors indicate the performance of the \ac{DA} relative to QIS3: green for better, yellow for equal, and blue for worse. A progress plot for \ac{DA}v2 is not possible due to the lack of the functionality to log intermediate improving cuts. The dots of instances G11, G48, G1, and G43 for \ac{DA}v2 are overlapping.}
	\label{fig:DAv3_vs_QIS3_progress}
\end{figure}

\Cref{fig:DAv3_vs_QIS3_progress} shows the progress of \ac{DA}v3's cut value relative to QIS3's over the runtime. For half of the instances, \ac{DA}v3 achieves an equivalent or better cut within the first few seconds. An equivalent figure for \ac{DA}v2 is not possible because it lacks the functionality to log intermediate improving cuts into a solution pool during execution. However, we also display in \cref{fig:DAv3_vs_QIS3_progress} the best cut achieved by \ac{DA}v2 (dots), as reported in \cref{table:QIS3vsDA}, along with the shortest runtime at which it was attained. These results are derived from multiple runs with time limits ranging from 1 to 10 seconds. For 6 out of 14 instances, \ac{DA}v2 found its best cut that matches or exceeds QIS3's cut in just a few seconds. For 4 instances, the cut was achieved in roughly half the time limit. This demonstrates the rapid convergence behavior of the \ac{DA} toward high-quality solutions \hyperlink{3a}{(3a)}.

\section{Conclusion}
\label{sec:conclusion}
We have conducted a comprehensive benchmarking study of Fujitsu’s second- and third-generation \acfp{DA}, \ac{DA}v2 and \ac{DA}v3, on the \ac{Max-Cut} problem. 
We compare their performance against selected best-performing \ac{MQLib} classical heuristics, D-Wave's hybrid quantum-classical solver (\ac{HS}), and a recently introduced quantum-inspired heuristic QIS3.

Our evaluation across 2,125 instances--spanning graph sizes from 200 up to approximately 53,000 variables--was based on the best objective value achieved within instance-specific time limits, ensuring as fair a comparison as possible. The results demonstrate that \ac{DA}v2 consistently found a better cut than the best classical heuristics on approximately 69.2\% of the benchmarked subset of instances, while \ac{DA}v3 maintains competitive performance with a success rate of about 60.8\%.  Notably, \ac{DA}v3's internal automatic scaling and rounding to integer coefficients proved suboptimal. By applying manual scaling, however, the success rate improved to approximately 69.1\%.
On 45 \ac{Max-Cut} instances selected by D-Wave, \ac{DA}v3 outperformed D-Wave's \ac{HS}, particularly on instances with integer weights. \ac{DA}v3 found lower-quality cuts for the majority of float-valued instances, though with a high relative cut ratio.
Finally, our comparison with QIS3 on 16 G-set instances reveals that one or both \ac{DA} versions achieve shorter runtimes and better solutions in most cases. However, for two large G-set instances, QIS3 shows a surprisingly good performance.
We further show that the \ac{DA} can have a rapid convergence behavior already before the time limits are met. 
Often, high-quality solutions can be reached within the first few seconds of runtime. 

Future work can expand this study in several directions. 
First, we aim to benchmark the \ac{DA} on constrained NP-hard problems to assess its effectiveness in more complex settings. Our results also suggested that performance is highly instance-dependent: for certain graph structures or sizes, the \ac{DA} quickly finds optimal or near-optimal cuts, while in other cases, traditional solvers or heuristics outperform it. This variation motivates a hybrid workflow that leverages different solvers depending on the instance features. Another promising direction is to incorporate preprocessing techniques into the \ac{DA}, tailored to a specific problem class such as the \ac{Max-Cut}, where it could further enhance its performance.

We emphasize that future benchmarking efforts would benefit from including modern parallelized or GPU-based classical solvers, such as multithreaded simulated annealing, parallel tempering, or GPU-accelerated heuristics. While our study is limited by the absence of such solvers, due to lack of publicly available and user-friendly implementations, we view their integration as a natural and important next step. We therefore recommend future works on such solvers to adopt and extend on our benchmarking design to complement and build upon our results presented here.

\section{Conflict of interest}
The \ac{DA} is a Fujitsu product, and Kirsch, Münch, Schinkel, and Walter work for the \emph{Fujitsu Germany GmbH}, which is a German Fujitsu subcompany. 
Their main work consists of consultations on quantum-inspired computing. 
Kliesch is the holder of the endowed professorship ``Quantum Inspired and Quantum Optimization'', which is financed by Fujitsu Services GmbH and the Dataport AöR.
By German laws, all university professorships are guaranteed scientific independence, and this requirement is also accommodated in the legal framework of this endowed professorship.

The authors declare that the comparison of the \ac{DA} with other methods was conducted to provide a fair and objective comparison, without giving special consideration to Fujitsu's interests. 
In fact, Winkler, Shaglel, and Kliesch initiated this work to independently determine the \ac{DA}'s performance without having to trust other sources. 
All authors made their best efforts to follow fair benchmarking standards as much as possible, given the technical feasibility. 

\section{Acknowledgment}
We thank John Silberholz for helpful discussions about methodology of Ref.~\cite{Dunning2018} and valuable comments regarding the \ac{MQLib} repository \cite{mqlib}.
We thank Mirko Arienzo, Nikolai Miklin, and Michel Krispin for discussions and valuable feedback during the development of this project.
Shaglel and Kliesch are funded 
by Fujitsu Germany GmbH and Dataport as part of the endowed professorship ``Quantum Inspired and Quantum Optimization''
and the Hamburg Quantum Computing project, which is co-financed by the ERDF of the European Union and the Fonds of the Hamburg Ministry of Science, Research, Equalities and Districts (BWFGB).

\section*{Appendices}
\pdfbookmark{Appendices}{Appendices}
\appendix
\section{\texorpdfstring{\ac{DA}v2}{DAv2} vs. best MQLib heuristics on x-Small--Small instances}
\label{appendix:small_instances_comparison}
We analyze the performance of \ac{DA}v2 in comparison with the corresponding best-performing \ac{MQLib} heuristics for x-Small and Small categories across all densities. As identified in \cref{table:category_heuristics} in \cref{subsec:best-of-37_heuristic}, the best heuristic for x-Small and Small Sparse instances is BURER2002, while for the Balanced and Dense instances it is PAL2004bMTS2. This analysis is omitted for \ac{DA}v3 due to often exceeding the time limit specified for instances in these categories by significantly more than 10\%, leading to an unfair comparison with other solvers (see \cref{appendix:runtime_deviation_analysis}).
\Cref{fig:xsmall_small_instances_visualizations} illustrates the size and density characteristics of the instances involved in this analysis, with a total of 1,306 instances. Most Balanced and Dense instances contain fewer than $\sim1,000$ vertices, and the majority of x-Small and Small instances are sparse, as shown by the nested inset in \cref{fig:xsmall_small_instances_hist} and the darker blue region of \cref{fig:xsmall_small_instances_scatter}.
\begin{figure}[t]
	\centering
 	\begin{subfigure}{0.485\textwidth}
 		\captionsetup{justification=raggedright, singlelinecheck=false, position=top}
 		\subcaption{}
		\includegraphics[width=\linewidth]{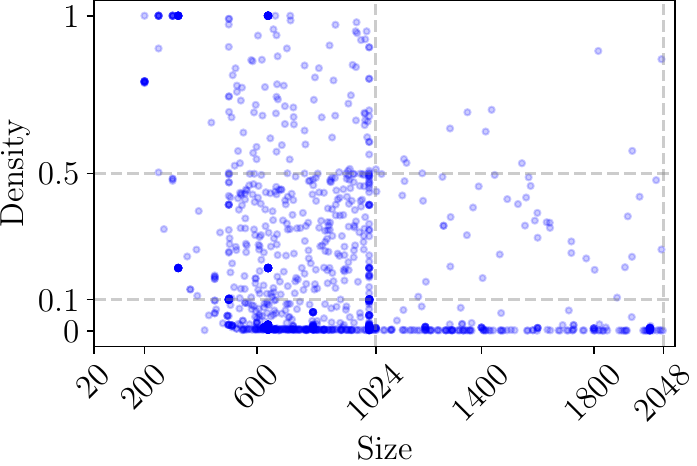}
		\label{fig:xsmall_small_instances_scatter}
 	\end{subfigure}
 	\begin{subfigure}{0.485\textwidth}	
 		\captionsetup{justification=raggedright, singlelinecheck=false, position=top}	
 		\subcaption{}
		\includegraphics[width=\linewidth]{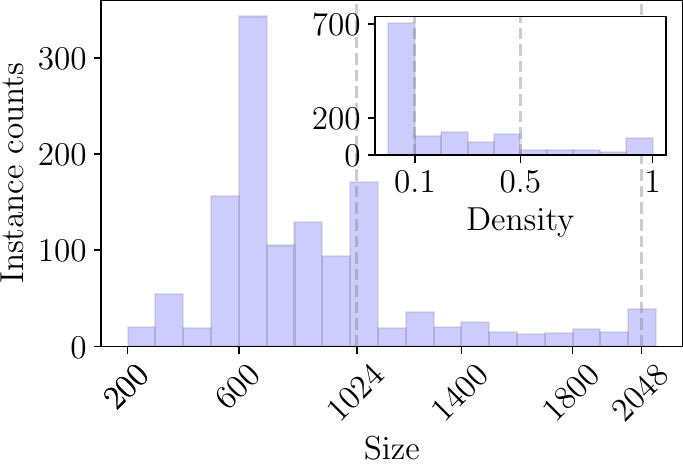}
		\label{fig:xsmall_small_instances_hist}
	\end{subfigure} 
	\caption{Distribution of instances by number of vertices and density for x-Small and Small category instances. The total number of instances is 1,306. (a) Scatter plot of instances in terms of their size and density. Darker blue regions indicate a higher concentration of instances at this size and density region. (b) Histogram showing instance counts by size range; the majority of instances are below $\sim1,000$ variables (in the x-Small category). Inset: Histogram showing instance counts by density range; the majority of instances are sparse.}
	\label{fig:xsmall_small_instances_visualizations}
\end{figure}

\begin{figure}
	\centering
	\includegraphics[width=\linewidth]{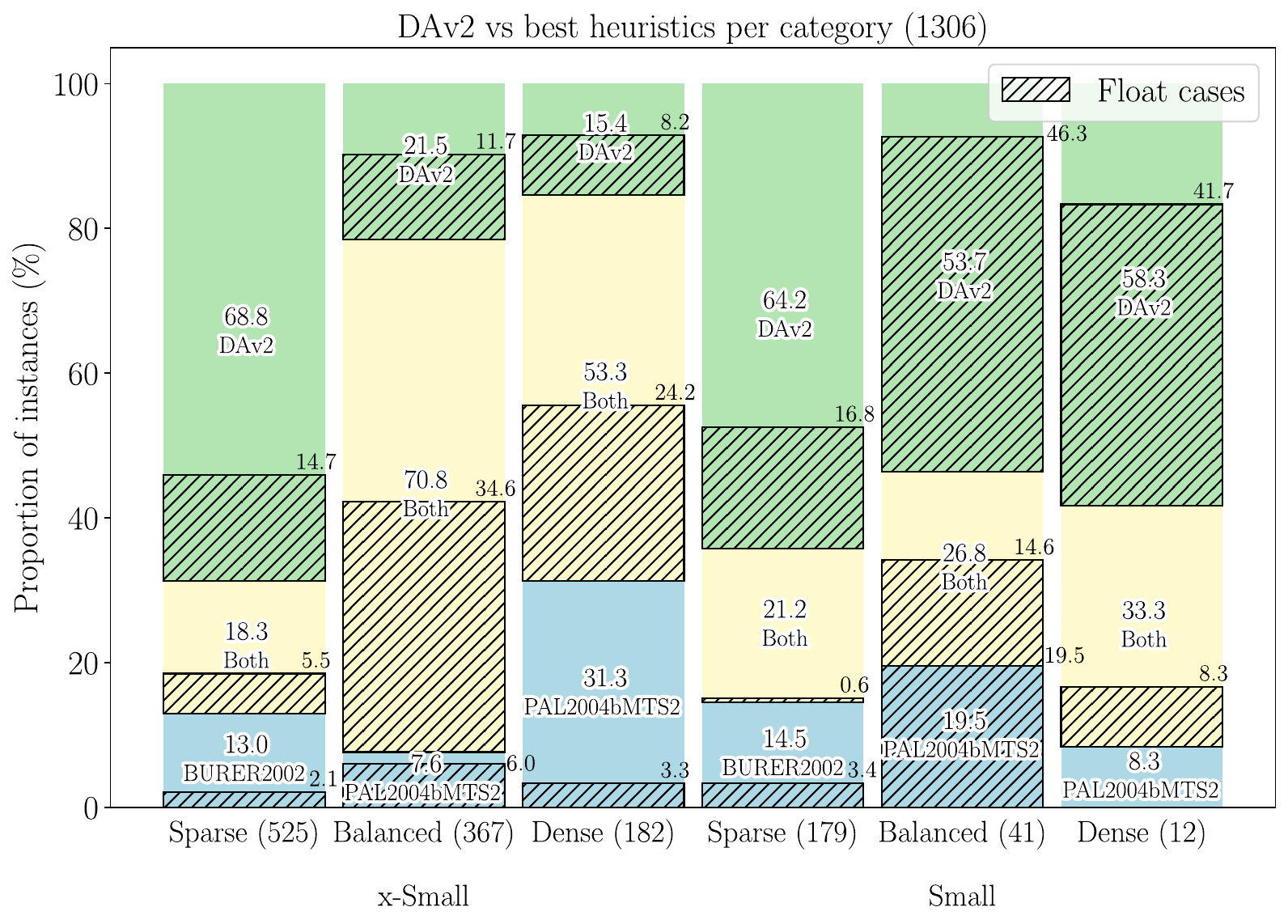}
	\caption{Comparison of the performance of \ac{DA}v2 with the best-performing \ac{MQLib} heuristic on x-Small and Small instances across all densities. Each bar represents the proportion of instances where \ac{DA}v2 yields a better cut (green), equal (yellow), or worse (blue) than the corresponding heuristic. Hatched bars indicate instances with floating-point cut values.}
	\label{fig:DAv2_xsmall_small}
\end{figure}

\Cref{fig:DAv2_xsmall_small} shows that on average, \ac{DA}v2 provides better solutions than BURER2002 in the corresponding categories: x-Small Sparse and Small Sparse (68.8\% and 64.2\%). PAL2004bMTS2 outperforms \ac{DA}v2 on x-Small Dense (31.3\% vs. 15.4\%). For a considerable number of instances, especially in x-Small Balanced and Dense, both find the same cut value.
On the other hand, \ac{DA}v2 yields better results than PAL2004bMTS2 on Small Balanced and Small Dense categories, finding a better cut in 53.7\% and 58.3\% of the instances, respectively. However, the number of instances in these categories is relatively small, so these results should be interpreted with caution.
 \begin{figure}
	\centering
	\includegraphics[width=\linewidth]{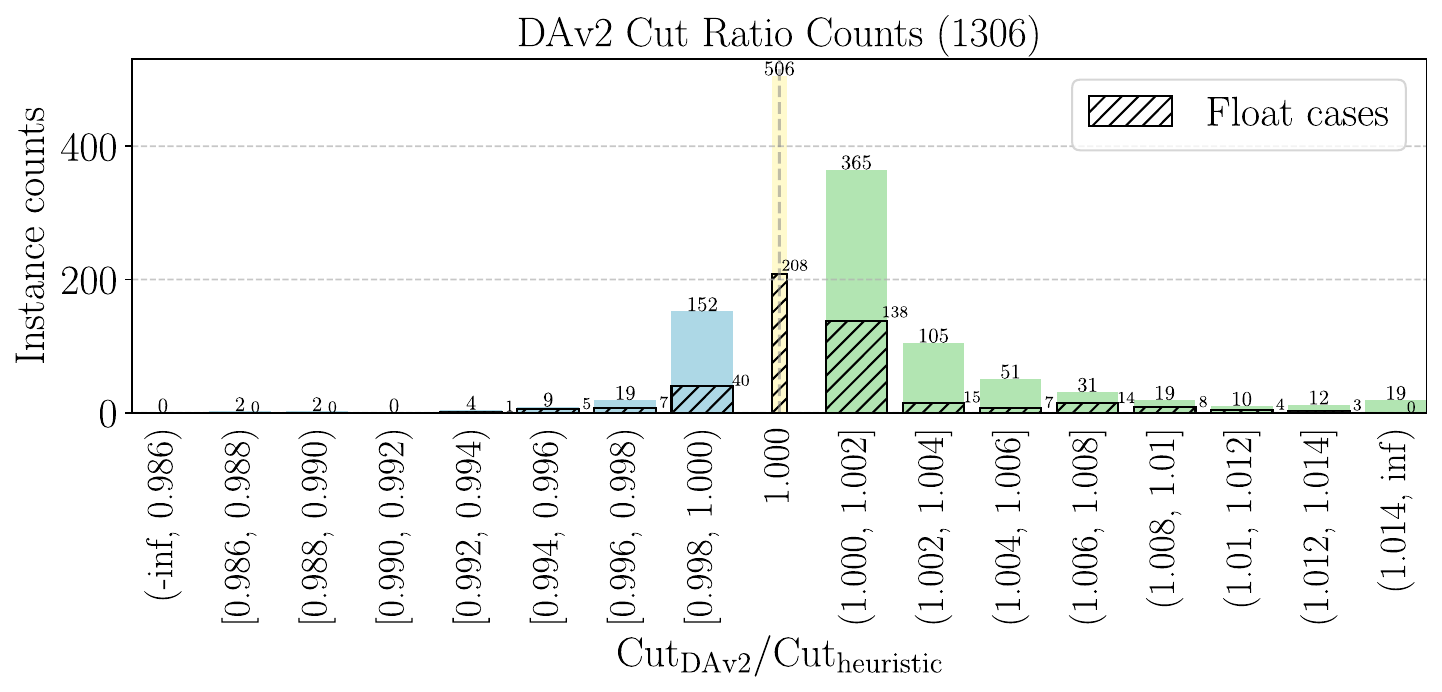}
	\caption{Distribution of all x-Small and Small instances over cut ratio ranges, where the ratio is defined as \ac{DA}v2 cut divided by that of the best-performing category heuristic. The `tie' bar at 1.0 (yellow) shows the count of instances when both solvers found an equal cut value. The dashed line at 1.0 separates `win' ranges (green bars) from `loss' ranges (blue bars). Hatched bars indicate instances with floating-point cut values.}
	\label{fig:DAv2_acc_counts_xsmall_small}
\end{figure}

\Cref{fig:DAv2_acc_counts_xsmall_small} presents a bar plot of instance counts across cut ratio ranges, defined as the ratio of the cut value achieved by \ac{DA}v2 to that achieved by the corresponding heuristic. The plot shows a clear skew to the values greater than 1.0 (i.e., higher green bars), highlighting an overall better performance of \ac{DA}v2 over the best-performing \ac{MQLib} heuristics on the x-Small and Small categories. Most `losses' of \ac{DA}v2 have high cut ratio between $[0.998,1.000)$. In total, \ac{DA}v2 achieves better cuts in 612 instances and equal in 506 out of 1306 instances, resulting in a `win' rate of $\sim46.86\%$, a `tie' rate of $\sim 38.74\%$, and a `loss' rate of $\sim 14.40 \%$.  As explained in the main text, we consider paired (by seed) run-level comparisons to assess statistical significance. On these categories, \ac{DA}v2 achieves better objective values in a majority ($\geq 3$) of five paired runs on 788 of 1306 instances ($\sim$ 60.34\%), which is highly significant according to a two-sided paired sign (binomial) test ($p = 8.03 \times 10^{-14}$). Runs in which \ac{DA}v2 and the corresponding category heuristic achieve identical objective values are conservatively counted in favor of the latter.

\section{Performance analysis on instances with known optimal solutions}
\label{appendix:optimal_solution_instances}

The performance of heuristic solvers is typically tested on large \ac{Max-Cut} instances, as such instances reveal scalability and robustness and better reflect real-world problem sizes, thereby enabling meaningful discrimination between strong and weak solvers. 
For large \ac{Max-Cut} instances, the optimal solution is usually unknown, which is 
standard in the field \cite{Vodeb2024, Dunning2018, biqmac2007, yang2025}. Consequently, solvers are commonly assessed using best-known cut values or through relative comparisons against established baseline heuristics, as done in this work.

To complement this evaluation, we additionally compare \ac{DA}v2 and \ac{DA}v3 against the best-performing heuristics identified in \cref{subsec:best-of-37_heuristic} on instances with known optimal solutions. Owing to the considerations outlined above, these (unstructured) instances are necessarily small.

We select multiple instance classes with known optimal solutions from BiqMac library \cite{biqmac2007}, which are also included in \ac{MQLib} \cite{mqlib}, as listed in \cref{table:biqmac-results}. Each instance class comprises multiple instances (indicated by the column
number) and varies in size and density, as reflected by $n$ and $d$ columns. The exact specifications of the instances are provided in library's documentation \cite{biqmac2007}.

For each solver, we report the number of instances per class for which the optimal solution is found within the instance-specific time limit (see \cref{subsec:baseline_timelimit}), as well as the worst accuracy, i.e.~ achieved cut/optimal cut, achieved across all instances in that class.
\begin{sidewaystable}
\centering
\small
\begin{tabular}{lllc|cc|cc|cc|cc|cc}
\toprule
\multirow{2}{*}{Instances} &
\multirow{2}{*}{$n$} &
\multirow{2}{*}{$d$} &
\multirow{2}{*}{$\#$} &
\multicolumn{2}{c|}{DAv2} &
\multicolumn{2}{c|}{DAv3} &
\multicolumn{2}{c|}{BURER2002} &
\multicolumn{2}{c|}{PAL2004bMTS2} &
\multicolumn{2}{c}{MERZ1999GLS} \\
\cmidrule(lr){5-6}
\cmidrule(lr){7-8}
\cmidrule(lr){9-10}
\cmidrule(lr){11-12}
\cmidrule(lr){13-14}
 & & & &
\# opt & worst acc.&
\# opt & worst  acc.&
\# opt & worst  acc.&
\# opt & worst  acc.&
\# opt & worst  acc.\\
\midrule

bqp & 100 & 0.1 & 10
 & 10 & 1
 & 10 & 1
 & 9 & 0.9973
 & 10 & 1
 & 10 & 1 \\

ising & 100--300 & 1 & 30
 & 0 & 0.9918
 & 6 & 0.9882
 & 1 & 0.9869
 & 3 & 0.9918
 & 4 & 0.9946 \\

w & 100 & 0.1,0.5,0.9 & 30
 & 29 & 0.9956
 & 30 & 1
 & 30 & 1
 & 30 & 1
 & 30 & 1 \\

t & 100--400 & 0.010025--0.04838 & 18
 & 9 & 0.9883
 & 13 & 0.9881
 & 8 & 0.9901
 & 2 & 0.9833
 & 9 & 0.9819 \\

pm & 80,100 & 0.1,0.99 & 40
 & 40 & 1
 & 40 & 1
 & 40 & 1
 & 40 & 1
 & 40 & 1 \\

gk & 20--200 & 0.0625--1 & 40
 & 40 & 1
 & 40 & 1
 & 35 & 0.8836
 & 40 & 1
 & 40 & 1 \\

g05 & 60,80,100 & 0.5 & 30
 & 30 & 1
 & 30 & 1
 & 30 & 1
 & 30 & 1
 & 30 & 1 \\

pw & 100 & 0.1,0.5,0.9 & 30
 & 30 & 1
 & 30 & 1
 & 28 & 0.9970
 & 30 & 1
 & 29 & 0.9975 \\

\bottomrule
\end{tabular}
\caption{Solver Performance comparison on instances with known optimal solutions from BiqMac Library \cite{biqmac2007} that exist also in \ac{MQLib} \cite{mqlib}.
Worst accuracy achieved by the solver across all instances in a given class is reported.}
\label{table:biqmac-results}
\end{sidewaystable}

For \texttt{bqp, pm, gk, g05,} and \texttt{pw} instance classes, nearly all solvers found the optimal solution for all instances. The main exceptions occur for BURER2002, which failed to reach optimality on one instance in \texttt{bqp} class (albeit with high accuracy), on five instances in \texttt{gk} class (with relatively low worst accuracy), and on two instances in \texttt{pw} class (with high worst accuracy).
For \texttt{w} instance class, all solvers found the optimal solution for all instances except \ac{DA}v2, which failed on a single instance but still achieved high accuracy.

The classes \texttt{ising} and \texttt{t} appear to be the most challenging. For \texttt{t} class, none of the solvers found the optimal solution for all instances; \ac{DA}v3 performed best, finding the optimum for 13 instances. Similarly, for the \texttt{ising} class, no solver achieved optimality across all instances. Here, \ac{DA}v3  again found the optimal solution for the largest number of instances, while \ac{DA}v2 did not for any instance.

All instances in these classes have integer weights that fit within 64-bit signed integer, so numerical rounding errors are unlikely to play a role. The results suggest that, particularly for the \texttt{ising} instances, the assigned time limit underestimates the problem hardness. Recall that \ac{DA}v3 often exceeds short imposed time limits (see \cref{appendix:DAv3_offset_selection} and \cref{appendix:runtime_deviation_analysis}), which may explain its superior performance on some instance classes.

\section{ Reliability of solvers measured by Coefficient of Variation (CV)}
\label{appendix:Solvers_stability}
\begin{figure}[!b]
		\centering

		\begin{subfigure}[b]{1\textwidth}
			\includegraphics[width=\textwidth]{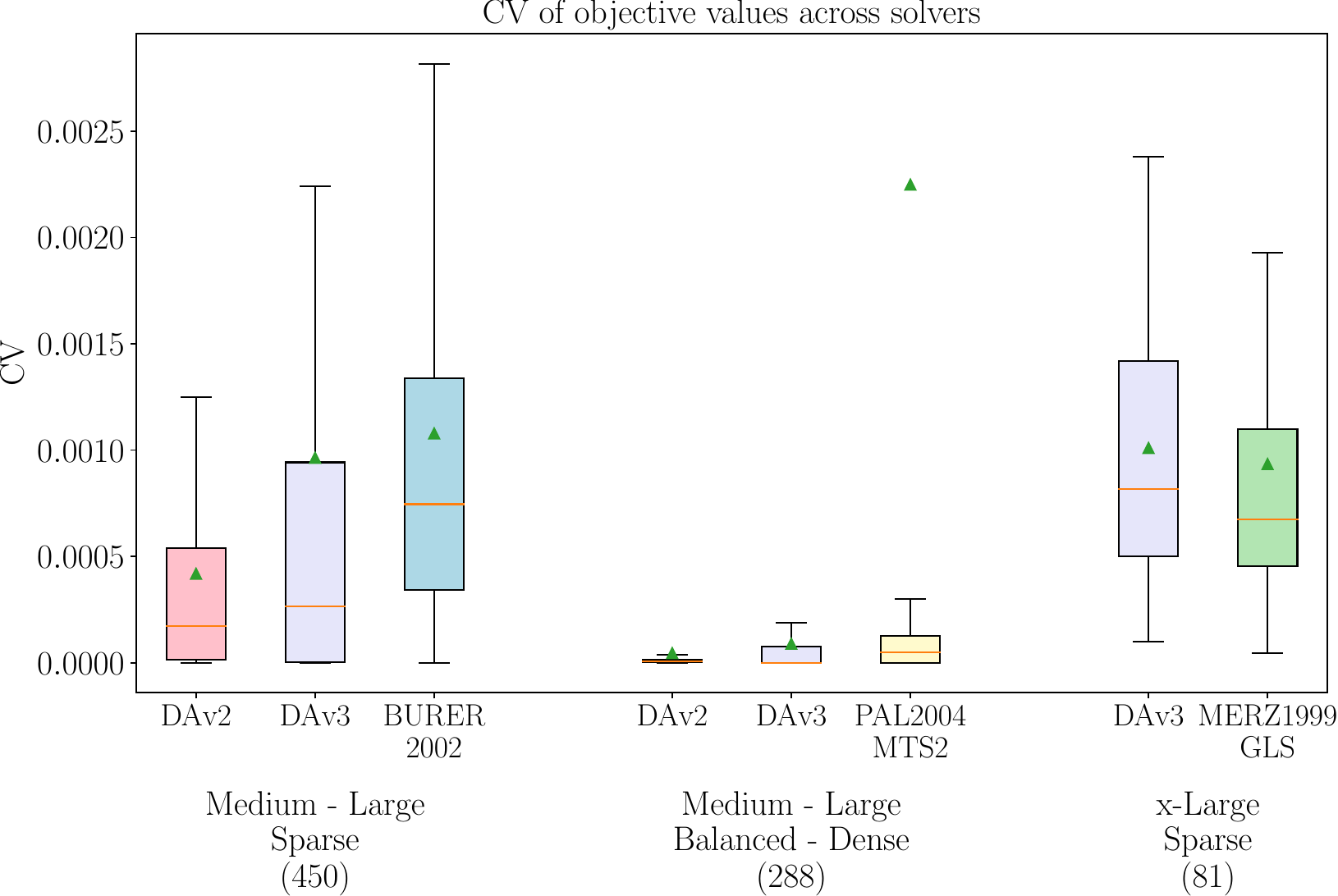}
			
		\end{subfigure}
        \\
        \vspace{2em}
		\begin{subfigure}[b]{1\textwidth}
			\includegraphics[width=\textwidth]{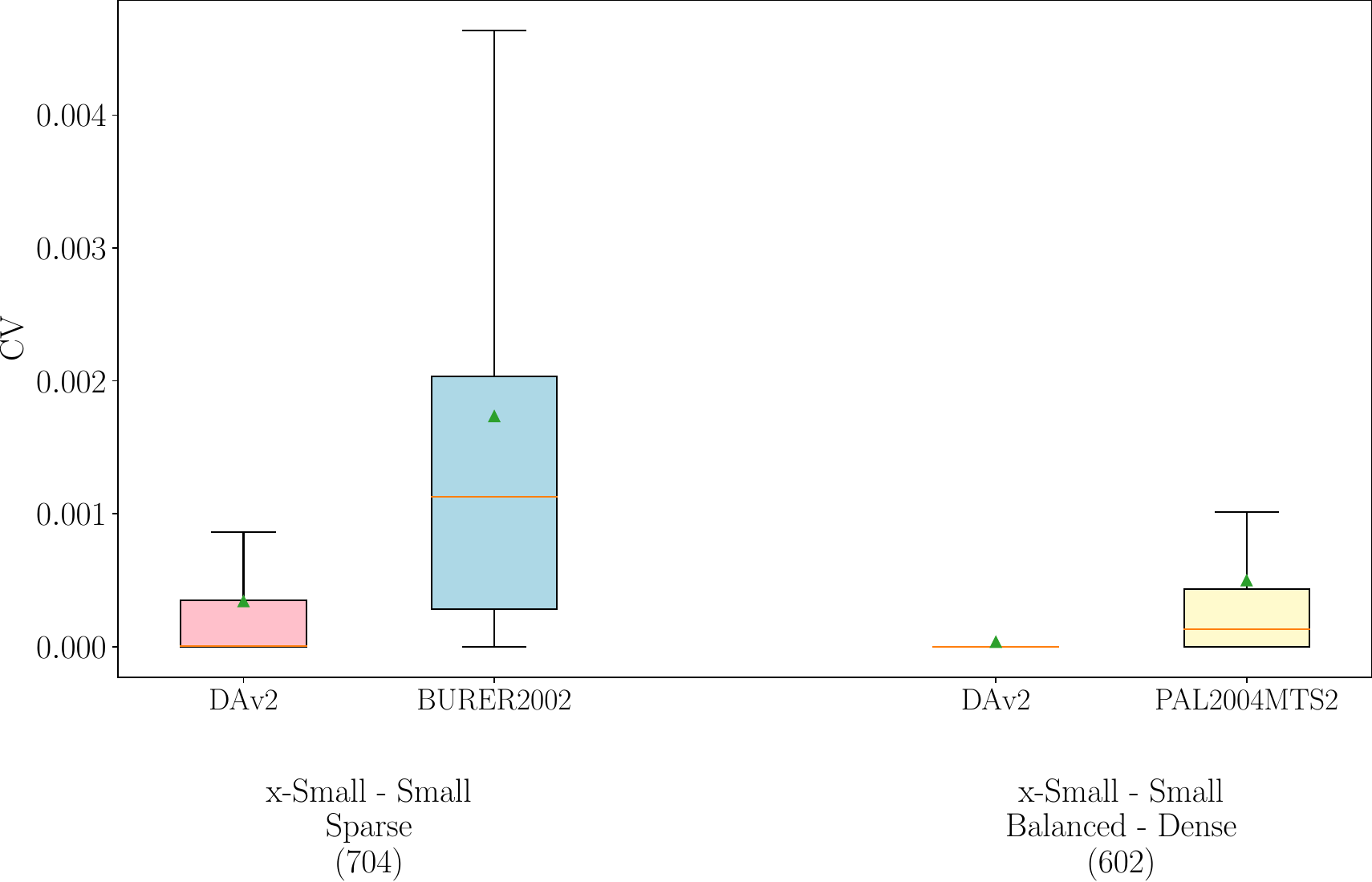}
		\end{subfigure}
        \caption{Box plots of the coefficient of variation (CV) of objective values across five independent runs, aggregated over instances in each size-density category studied in \cref{subsec:DA_vs_mqlib_heuristics} (above) and in \cref{appendix:small_instances_comparison} (below). The number of instances per category indicated in parentheses. Lower \ac{CV} values indicate lower run-to-run 
        variability of the solver's solution. The mean value is indicated by a green triangle and the median by an orange horizontal line. For clarity of presentation, outliers are not displayed.}
        \label{fig:CV_obj}
    \end{figure}

In the main text, we reported performance comparisons between the \ac{DA} and the baseline solvers using the best objective value obtained over 5 independent runs with 5 different seeds per instance. This choice was motivated by an application-oriented setting in which practitioners seek the best solution obtained from a small number of independent trials.
To further show that this choice does not bias the reported conclusions -- and in addition to the statistical inference assessments that are independent of the best-of-5 metric -- we explicitly analyze the run-to-run variability by presenting distribution of the \ac{CV} across the 5 runs objective values for each instance. The \ac{CV} is defined as
\begin{equation}
   \mathrm{CV} = \frac{\mathrm { Standard \, deviation}}{\mathrm {Mean} },
\end{equation}
which provides a scale-normalized and dimensionless measure of run-to-run variability.
\cref{fig:CV_obj} shows these \ac{CV} distributions as box plots grouped by instance category and solver. The upper plot corresponds to the study on Medium to x-Large instances presented in \cref{subsec:DA_vs_mqlib_heuristics}, while the lower plot corresponds to the x-Small to Small instances study discussed in \cref{appendix:small_instances_comparison}.

Overall, \cref{fig:CV_obj} shows that \ac{CV} values are very small across all solvers and categories, indicating a high degree of robustness and limited stochastic variability in the obtained objective values. Across all applicable categories, \ac{DA}v2 exhibits the lowest variability, as reflected by the lowest mean (green triangle) and median (orange horizontal line) \ac{CV} values. \ac{DA}v3 is generally more reliable than BURER2002 and PAL2004bMTS2 heuristics, but less reliable than MERZ1999GLS. Notably, PAL2004bMTS2 produces several extreme outliers for the Medium - Large and Balanced - Dense instances, which is reflected by the high \ac{CV} mean value. 
We speculate that for larger and denser graphs the outcome concentrate more sharply around the typical behaviour. For too easy graphs, the behaviour also concentrates at the optimal solution.
	\begin{figure}[!h]			
		\begin{subfigure}[b]{1\textwidth}
			\includegraphics[width=\textwidth]{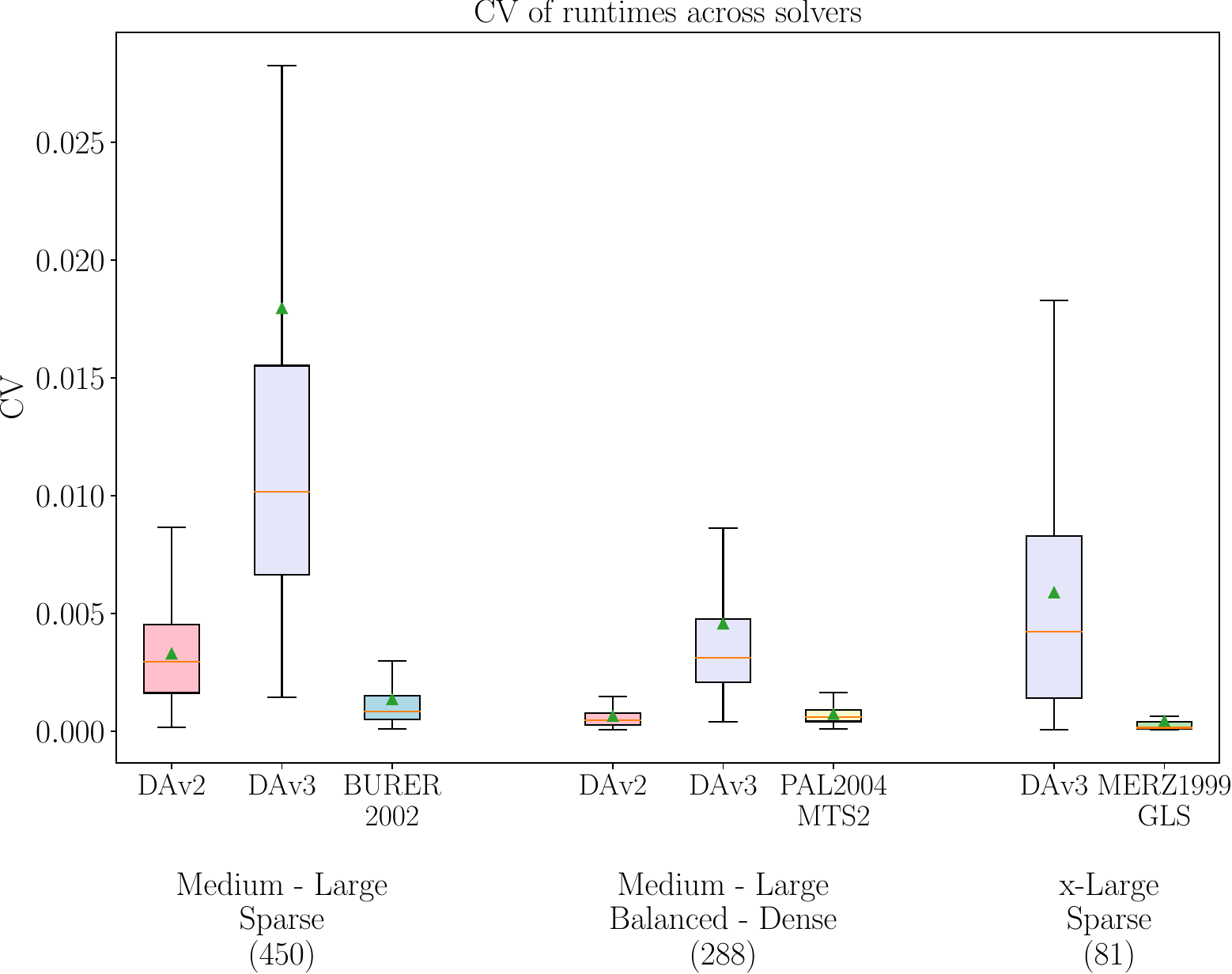}
		\end{subfigure}
        \vspace{2em}
        \\
		\begin{subfigure}[b]{1\textwidth}
			\includegraphics[width=\textwidth]{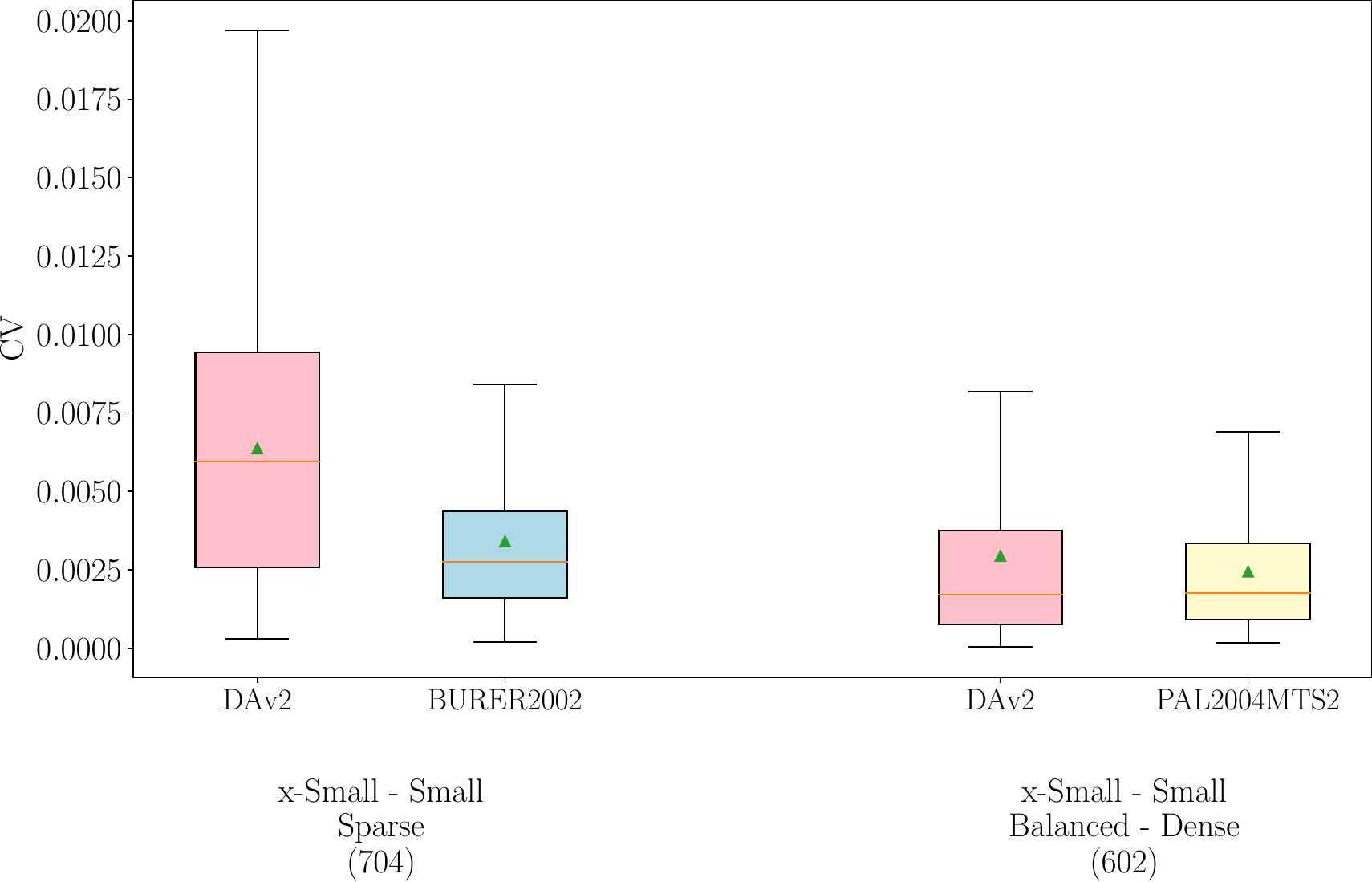}
		\end{subfigure}
        \caption{Box plots of the coefficient of variation (CV) of runtimes across five independent runs, aggregated over instances in each size-density category studied in \cref{subsec:DA_vs_mqlib_heuristics} (above) and in \cref{appendix:small_instances_comparison} (below). The number of instances per category indicated in parentheses. Lower \ac{CV} values indicate lower run-to-run variability of the solver's runtimes. The mean value is indicated by a green triangle and the median by an orange horizontal line. For clarity of presentation, outliers are not displayed.}
        \label{fig:CV_runtime}
    \end{figure}
\clearpage

\Cref{fig:CV_runtime} above further reports the distributions of \ac{CV} of the runtime across the 5 runs per instance. The \ac{MQLib} category heuristics generally exhibit lower run-to-run runtime variability than both \ac{DA} v2 and v3 across almost all instance categories. \ac{DA}v2 is clearly more reliable than \ac{DA}v3. Nevertheless, the overall \ac{CV} values remain small for all solvers and categories.

\section{DAv2 runtime estimation}
\label{appendix:DAv2_runtime_fitting}
To allow for \ac{DA}v2 to terminate within a specific time range, for the purpose explained in \cref{subsec:baseline_timelimit}, we have obtained fitted functions to estimate the annealing time and CPU time based on the input number of variables, runs, and iterations. 
To obtain these estimates, we generated runtime data of \ac{DA}v2 across all instances using various combinations of run and iteration numbers. The number of runs was varied from minimum to maximum in steps of 16, i.e.~${16, 32, 48, 64, 80, 96, 112, 128}$. The number of iterations 
was set in relation to the number of variables as $f \times n^2$ where $f \in \{1,2,4\}$.

The runtime strongly depends on the size category introduced in \cref{subsec:Selection_of_relevant_instances} to which a given instance belongs. 
Accordingly, the annealing time and CPU time functions are defined as 
\begin{align}
&\mathrm{Annealing\_time(runs, iterations)} = a (\mathrm{runs} \times \mathrm{iterations}) + b \label{eq:Anneal_time}\\ 
&\mathrm{CPU\_time(runs, } n ) = c ( n^2 \times \mathrm{runs})  + d (n \times \mathrm{runs}) + e n^2 \nonumber \\
& \qquad\qquad\qquad\qquad\qquad+ k (\mathrm{runs})+ g n+ h \label{eq:CPU_time}
\end{align}
where the fitting parameters $a, b, c, d, e, k, g, h,$ are given in \cref{table:parameters_of_fitting_functions} for each size category. Both fitted functions return time in milliseconds; time limits specified in seconds are converted to milliseconds before applying the estimates.
We used these equations to estimate the number of runs and iterations required for each instance to as much as possible meet a specified time limit. Specifically, we aimed to select the largest combination of run and iteration numbers that remains within 110\% of the time limit.
\begin{table}[!t]
\centering
\begin{tabular}{|c||c|c|}
	\hline & x-Small: $20 \leq n < 1024$  &  Small: $ 1024 \leq n < 2048$ \\ 
	\hline\hline \label{Table:fitted_parameters}
	$a $ & 2.0081 $\times 10^{-6}$ & 2.0017 $\times 10^{-6}$  \\ 
	$b$ & 13.2942 & 48.6364 \\
	$c$ & -0.5576 $\times 10^{-7}$  & 6.2894 $\times 10^{-7}$  \\ 
	$d$ & 0.0007 & -0.0007 \\ 
	$e$ & 2.9877 $\times 10^{-6}$  & 3.5768 $\times 10^{-6}$ \\
	$k$ & -0.0101 & 0.7396 \\ 
	$g$ & -0.0020 & 0.0056 \\ 
	$h$ & 4.3422 & 5.3949 \\
	\hline & Medium: $ 2048 \leq n < 4096$ & Large: $ 4096 \leq n < 8192$ \\
	\hline 
	$a $ & 2.0010 $\times 10^{-6}$  & 2.0005 $\times 10^{-6}$  \\
	$b$ & 193.8656 & 126.2170 \\
	$c $ & 7.8667 $\times 10^{-7}$  & 7.4802 $\times 10^{-7}$  \\ 
	$d$ & -0.0015 & -0.0002 \\
	$e$ & 4.1800 $\times 10^{-6}$  & -9.6817 $\times 10^{-6}$  \\ 
	$k$ & 1.8767 & -3.7253 \\
	$g$ & -0.0056 & 0.1548 \\ 
	$h$ & 59.8780 & -264.9900  \\
	\hline
\end{tabular}
\caption{Parameters of \ac{DA}v2 annealing and CPU time fitted functions, \cref{eq:Anneal_time,eq:CPU_time}, respectively. }
\label{table:parameters_of_fitting_functions}
\end{table}

The procedure is as follows: we begin with the minimum values--16 runs and $\max(10,000, n^2)$ iterations. We first estimate the CPU time under these values using \cref{eq:CPU_time}. From 99\% of the time limit, we subtract the estimated CPU time along with other fixed overheads, such as sampling\footnote{The time for setting the start and end temperature of the annealing process on \ac{DA}v2. We apply the sampling procedure described in \cite{chris2023} to achieve flip probability of 99\% in the beginning and 1\% after half of the annealing time.} time and scaling\footnote{Rescaling the \ac{QUBO} matrix by multiplying it with a factor. This step is essential for converting floating-point elements to integers and subsequently adjusting the objective value.} time to determine the remaining time available for annealing. The 99\% threshold allows for a buffer, and the sampling and scaling times are instance-specific constants independent of the number of runs and iterations and are extracted from the collected runtime data.
Based on this, we estimate the possible number of iterations using Eq.~\eqref{eq:Anneal_time}. 
If this value is more than the previous number of iterations, we update the iteration number accordingly. If it exceeds 40,000 iterations and the estimated total time (CPU time + annealing time + fixed overheads) is still below 99\% of the time limit, higher runs are explored. Otherwise, higher runs are not considered further.
\begin{figure}[!t]
    \centering
    \includegraphics[width=\linewidth]{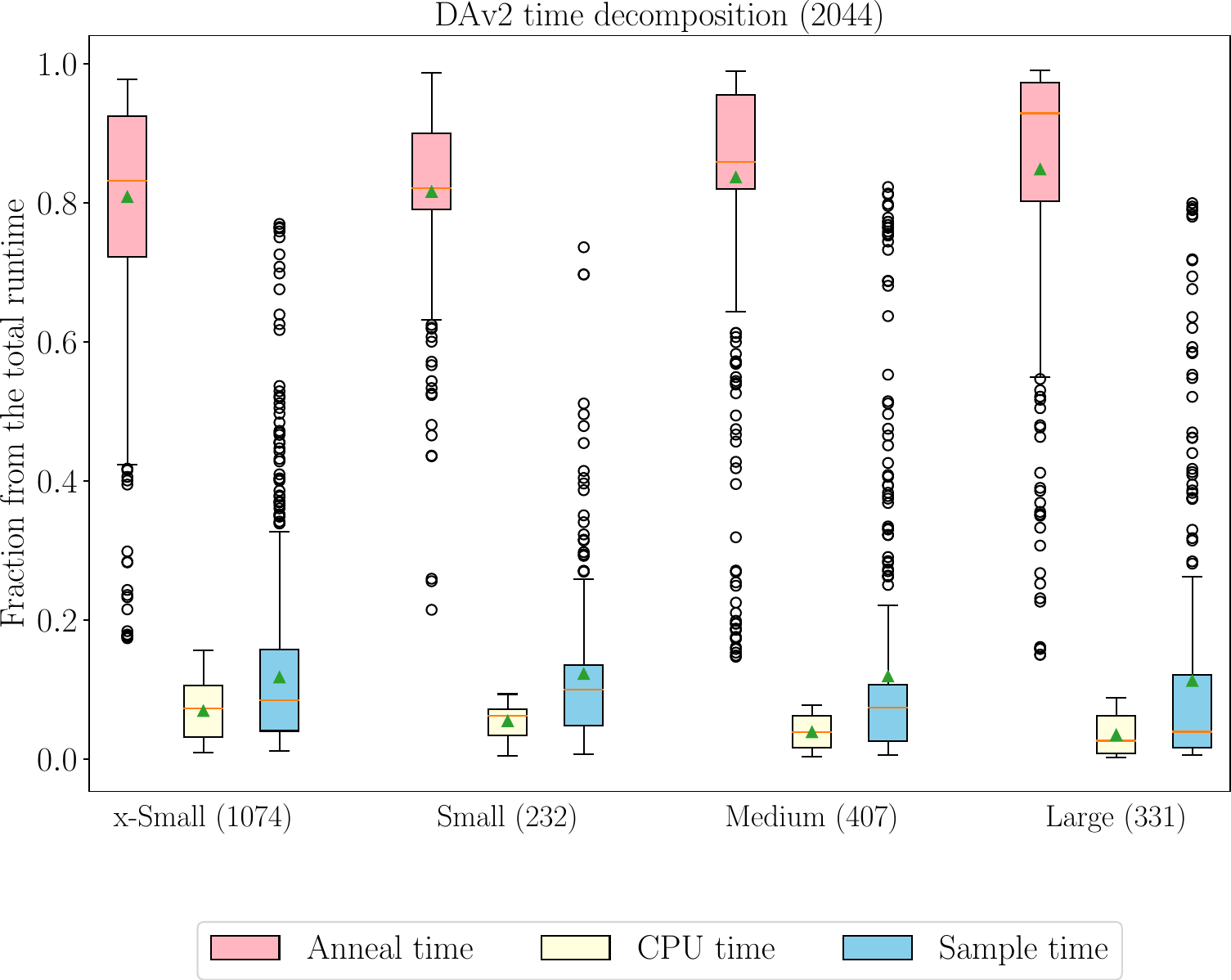}
    \caption{Distribution of relative contributions of \ac{DA}v2 primary runtime components: anneal, CPU, and sample times, to the total runtime across 2,044 instances. The mean value is indicated by a green triangle and the median by an orange horizontal line.}
    \label{fig:DAv2_time_decomposition}
\end{figure}

To illustrate the distribution of the relative contributions of \ac{DA}v2's primary runtime components to the total runtime, we present \cref{fig:DAv2_time_decomposition} on 2,044 instances, grouped by instance size categories. The three primary components are anneal time, CPU time, and sample time. Recall that throughout this study, we report the total runtime, which includes these three components as well as additional overheads discussed earlier. As shown in the figure, anneal time typically constitutes the largest fraction of the total runtime. Some outliers in the low anneal-time fraction arise because the annealing time is allocated as the remaining time after accounting for the other time components. This pattern is consistent across all instance size categories and remains stable for each of the three runtime components.

\section{DAv3 time limit offset determination}
\label{appendix:DAv3_offset_selection}
As noted in \cref{subsec:baseline_timelimit}, \ac{DA}v3 is more prone to far exceed shorter time limits, so our focus is on identifying an appropriate time offset specifically for such cases. Due to its stochastic runtime behavior, we were unable to obtain a fitted runtime function as done for \ac{DA}v2 in \cref{appendix:DAv2_runtime_fitting}.
Instead, we determine a suitable time limit offset empirically. To this end, we executed \ac{DA}v3 on all Medium to x-Large instances that have been assigned by the algorithm in \cref{subsec:baseline_timelimit} time limits between 0.25 and 100 seconds,
with an offset-adjusted time limit defined in \cref{eq:time_offset} with various offsets: 0 to 5 seconds in 1-second increments. Our goal is to identify the offset that minimizes the number of instances exceeding the 10\% time safety margin while achieving the highest possible average cut ratio.
The average cut ratio for each offset is computed as
\begin{equation}
	\mathrm{Average\_cut\_ratio} (\%) := \frac{1}{N} \left(\sum_{i=1} ^N \frac{\mathrm{cut}_i}{\mathrm{best}_i} \right) \times 100
\end{equation}
where $\mathrm{cut}_i$ is the achieved objective value for instance $i$ for each corresponding offset, $\mathrm{best}_i$ is the best objective value found across all available offset runs for instance $i$, and $N = 819$ is the total number of instances in the Medium to x-Large categories.
\begin{figure}[t]
	\centering
	\includegraphics[width=\linewidth]{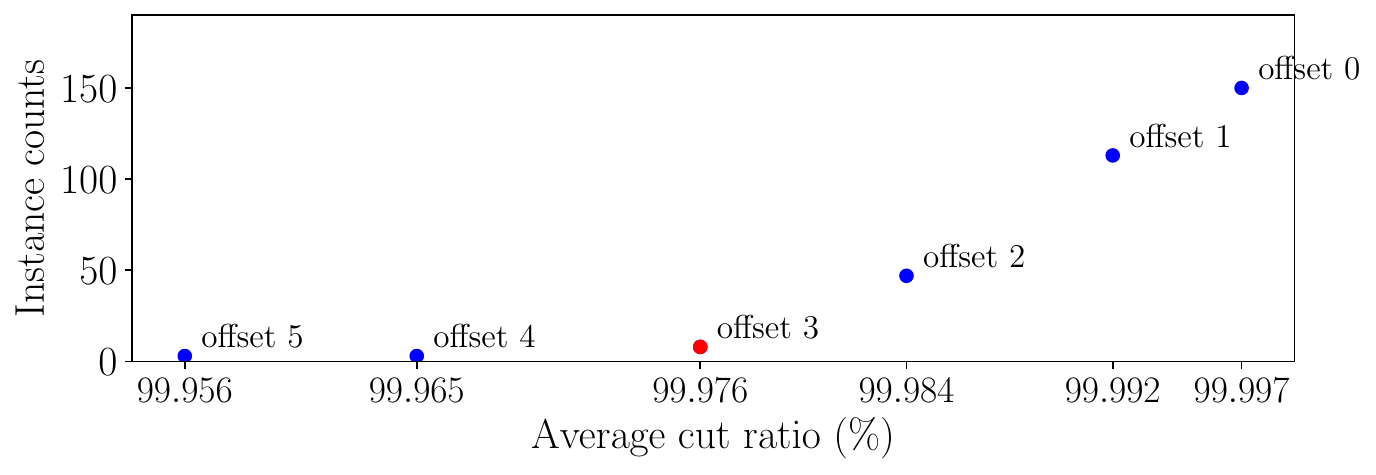}

	\begin{minipage}{\textwidth}
		\centering
		\resizebox{\textwidth}{!}{
		\begin{tabular}{|c|c|c|c|c|c|c|}
			\hline
			\textbf{Offset} (seconds) & 0 & 1 & 2 & 3 & 4 & 5 \\
			\hline
			\ \makecell{Instance counts with \\ runtime > 110\% of limit} & 150 & 113 & 47 & 8 & 3 & 3 \\
			\hline 
			Average cut ratio (\%) & 99.9971 & 99.9916 & 99.9838 & 99.9760 & 99.9645 & 99.9560 \\ 
			\hline
		\end{tabular}}
	\end{minipage}
	
	\caption{Trade-off between average cut ratio and time-limit adherence across different offset values for \ac{DA}v3. An offset of 3 seconds (red point) offers the best balance between runtime compliance and cut ratio quality.}
	\label{fig:time_accuracy_plot}
\end{figure}
In \cref{fig:time_accuracy_plot}, we show how the offset affects the average cut ratio and the number of instances whose runtime exceeds 110\% of the time limit. An offset of 3 seconds fairly reduces the number of instances that violate the safety margin, with only a marginal decrease in average cut ratio. Higher offsets yield minimal improvements but come with additional quality losses, making them less favorable.

Unfortunately, an explicit decomposition of \ac{DA}v3's runtime into different time components is not accessible; therefore, a plot analogous to \cref{fig:DAv2_time_decomposition} for \ac{DA}v2 is not possible.

\section{Runtime deviation of solvers from baseline time limit}
\label{appendix:runtime_deviation_analysis}
\begin{figure}[H]
		\centering

		\begin{subfigure}[b]{0.49\textwidth}
			\includegraphics[width=\textwidth]{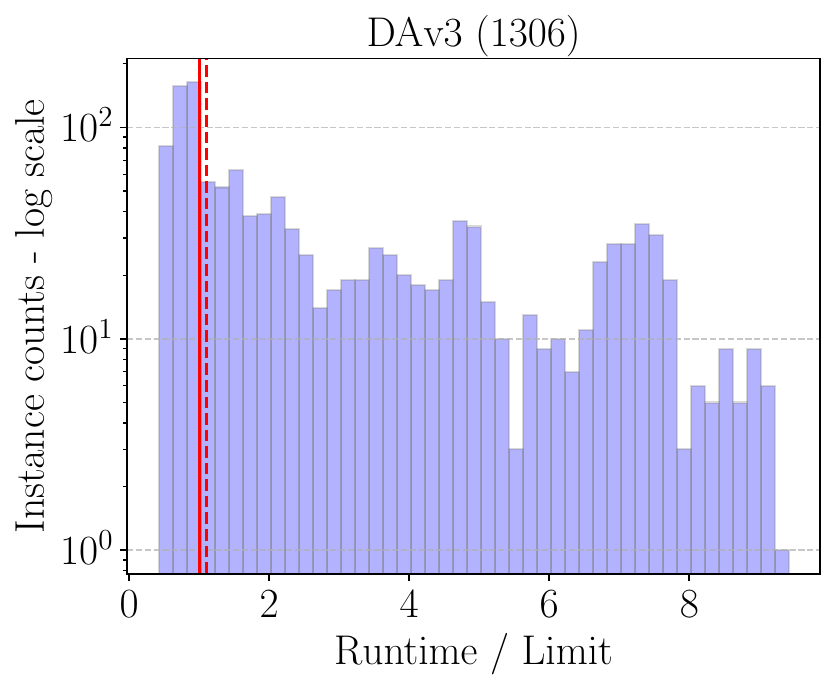}
			\caption{x-Small to Small categories (not analyzed).}
			\label{fig:time_analysis_DAv3_small}
		\end{subfigure}
		\begin{subfigure}[b]{0.49\textwidth}
			\includegraphics[width=\textwidth]{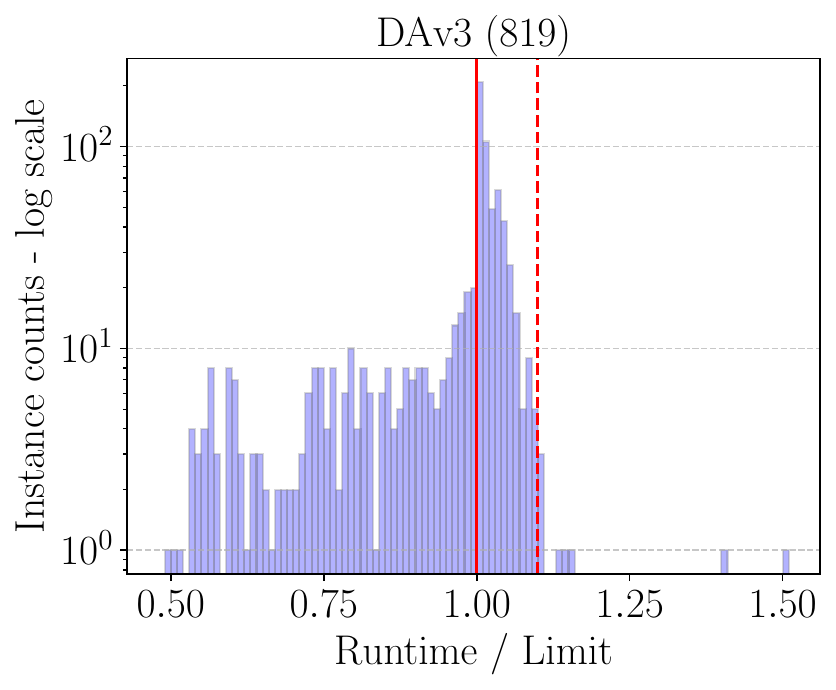}
			\caption{Medium to x-Large categories. \\ \vspace{1.3em} }
			\label{fig:time_analysis_DAv3}
		\end{subfigure}

		\vspace{0.4cm}

		\begin{subfigure}[b]{0.49\textwidth}
			\includegraphics[width=\textwidth]{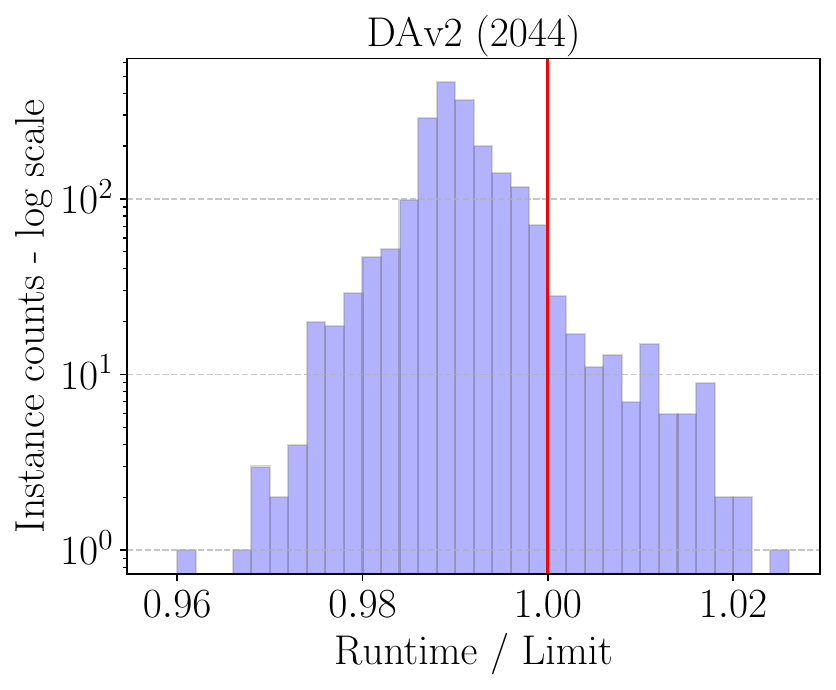}
			\caption{x-Small to Large categories.}
			\label{fig:time_analysis_DAv2}
		\end{subfigure}
		\begin{subfigure}[b]{0.49\textwidth}
			\includegraphics[width=\textwidth]{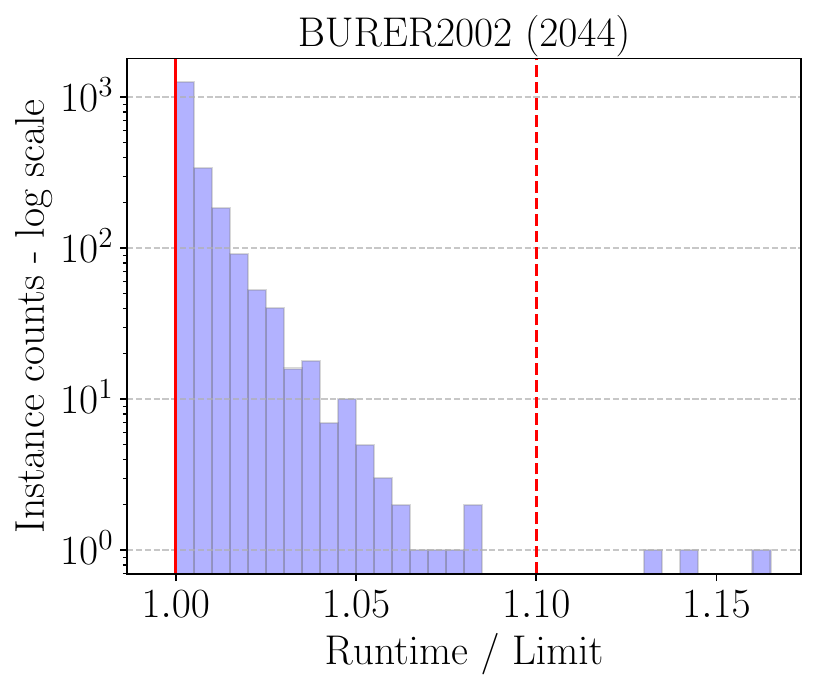}
			\caption{x-Small to Large categories.}
			\label{fig:time_analysis_Burer}
		\end{subfigure}
		
		\vspace{0.55cm}  
		
		\begin{subfigure}[b]{0.49\textwidth}
			\includegraphics[width=\textwidth]{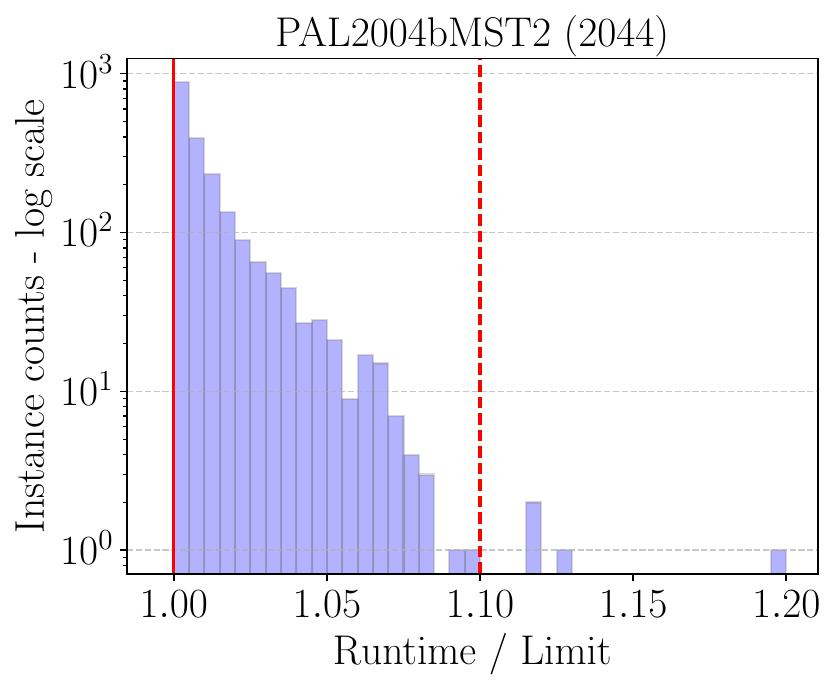}
			\caption{x-Small to Large categories.}
			\label{fig:time_analysis_Pal}
		\end{subfigure}
		\begin{subfigure}[b]{0.49\textwidth}
			\includegraphics[width=\textwidth]{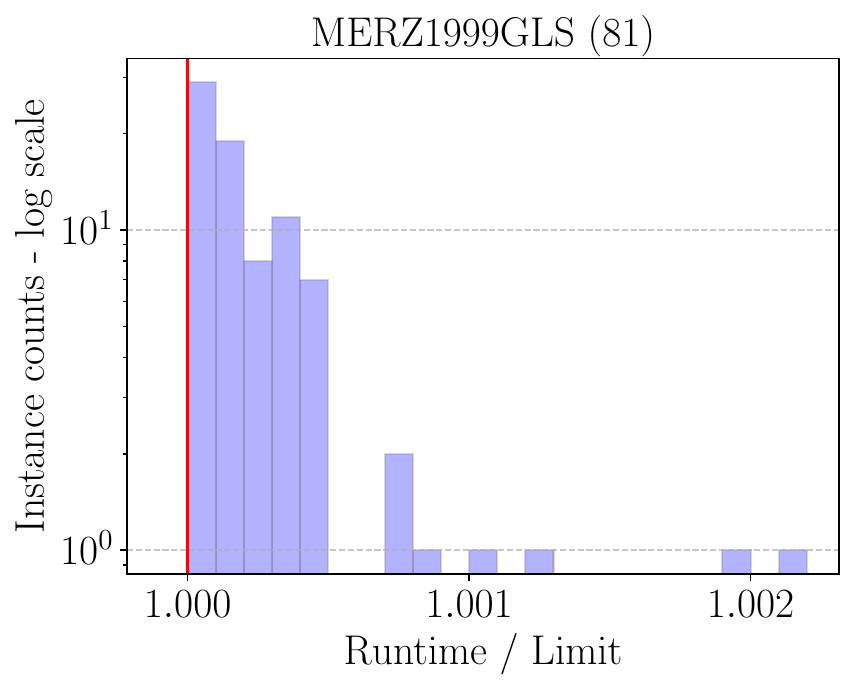}
			\caption{x-Large category.}
			\label{fig:time_analysis_Merz}
		\end{subfigure}
		
		\caption{Histogram (log scale) of relevant instance counts distributed over solver runtime-to-limit ratio. The solid red line at 1 indicates the point where runtime equals the time limit, while the dashed red line marks a $10\%$ exceedance margin that we enforced for our solvers on the majority of instances. \ac{DA}v2 and MERZ1999GLS remain within this margin, whereas \ac{DA}v3 in (b), BURER2002, and PAL2004bMTS2 exceed it for very few instances. Panel (a) illustrates why the corresponding \ac{DA}v3 results were excluded from the analysis.}
		\label{fig:time_analysis}
	\end{figure}

Solvers do not typically terminate exactly at the specified time limit. To account for this, we define a safety margin with an upper bound of $10\%$. In \cref{fig:time_analysis}, we present histograms showing instance counts by each solver's runtime-to-limit ratio. 
As shown in \cref{fig:time_analysis_DAv2}, the runtime fitting of \ac{DA}v2, detailed in \cref{appendix:DAv2_runtime_fitting}, was largely successful.
However, it also occasionally resulted in a runtime shorter than the assigned limit. 
It still exceeds the time limit slightly in some cases, but it is still well below the $10\%$ upper limit of the safety margin.

Although \ac{DA}v3 includes a time limit termination feature and an added buffer, it still exceeds the safety margin on a significant number of instances--especially those in the x-Small and Small categories (see \cref{fig:time_analysis_DAv3_small}). 
This is partially due to a 1-second minimum runtime, as well as the communication overhead between the \ac{DAU} and the software layer, making it more likely to exceed the specified time limit.
As a result, we restrict our analysis for \ac{DA}v3 to larger instances; the raw data are nevertheless provided in the accompanying GitHub repository \cite{shaglel2025dabenchmark}. The time analysis for \ac{DA}v3 is presented in \cref{fig:time_analysis_DAv3} for the Medium through x-Large categories. For a large number of instances, the runtime of \ac{DA}v3 is far below the time limit--more so than \ac{DA}v2--with runtimes dropping to as low as $50\%$ of the limit. This could partly explain why \ac{DA}v2 provides a higher `win' ratio than \ac{DA}v3.

As expected, \ac{MQLib} solvers also often exceed the time limit as shown in \cref{fig:time_analysis_Burer,fig:time_analysis_Pal,fig:time_analysis_Merz}. 
In particular, BURER2002 and PAL2004bMTS2 exceeded the safety margin for multiple instances. Overall, the \ac{DA} runtimes remain on average lower than those of the \ac{MQLib} heuristics.

\newpage

\section*{Acronyms}
\begin{acronym}[POVM]
	\acro{AC}{acronym}
	\acro{AGF}{average gate fidelity}
	
	\acro{BOG}{binned outcome generation}
	
  \acro{CV}{coefficient of variation}
	\acro{CP}{completely positive}
	\acro{CPT}{completely positive and trace preserving}
	\acro{CS}{compressed sensing} 
	
	\acro{DAU}{Digital Annealer Unit}
	
	\acro{GST}{gate set tomography}
	\acro{GUE}{Gaussian unitary ensemble}
	
	\acro{HOG}{heavy outcome generation}
	\acro{HS}{hybrid solver}
	
	\acro{MBL}{many-body localization}
	\acro{ML}{machine learning}
	\acro{MLE}{maximum likelihood estimation}
	\acro{MPO}{matrix product operator}
	\acro{MPS}{matrix product state}
	\acro{MUBs}{mutually unbiased bases} 
	\acro{MW}{micro wave}
	\acro{Max-Cut}{maximum cut}
	\acro{MQLib}{Max-Cut and QUBO instances library}

	\acro{NISQ}{noisy and intermediate scale quantum}
	
	\acro{POVM}{positive operator valued measure}
	\acro{PVM}{projector-valued measure}
	
	\acro{QAOA}{quantum approximate optimization algorithm}
	\acro{QML}{quantum machine learning}
	\acro{QMT}{measurement tomography}
	\acro{QPT}{quantum process tomography}
	\acro{QUBO}{quadratic unconstrained binary optimization}
	
	\acro{RDM}{reduced density matrix}
	
	\acro{SDP}{semidefinite programming}
	\acro{SFE}{shadow fidelity estimation}
	\acro{SIC}{symmetric, informationally complete}
	\acro{SPAM}{state preparation and measurement}
	
	\acro{RB}{randomized benchmarking}
	\acro{rf}{radio frequency}
	
	\acro{TT}{tensor train}
	\acro{TV}{total variation}
	
	\acro{VQA}{variational quantum algorithm}
	
	\acro{VQE}{variational quantum eigensolver}
	\acro{VQA}{variational quantum algorithm}
	
	\acro{XEB}{cross-entropy benchmarking}

	\acro{DA}{Digital Annealer}
	\acro{ASIC}{application-specific integrated circuit}
	\acro{CMOS}{complementary metal-oxide-semiconductor}
	\acro{Biq Mac}{binary quadratic and Max-Cut}

\end{acronym}

\bibliographystyle{myapsrev4-2} 
\bibliography{mybib,mk} 
\end{document}